\documentclass[prb, twocolumn, superscriptaddress]{revtex4}
\usepackage{graphicx}
\usepackage{dcolumn}
\usepackage{amsmath,amsbsy,amssymb,graphics}
\usepackage{bm}
\usepackage{mathrsfs,color}

\usepackage{nicefrac}

\usepackage{xcolor}
\usepackage{amsmath}
\usepackage{epsfig}
\usepackage{dcolumn}
\usepackage{amssymb}

\begin{document}

\title{Local interpretation of time-resolved X-ray absorption in Mott insulators: \\ Insights from nonequilibrium dynamical mean-field theory}

\author{Philipp Werner}
\affiliation{Department of Physics, University of Fribourg, 1700 Fribourg, Switzerland}
\author{Denis Golez}
\affiliation{Jozef Stefan Institute, Jamova 39, SI-1000 Ljubljana, Slovenia}
\affiliation{Faculty of Mathematics and Physics, University of Ljubljana, Jadranska 19, 1000 Ljubljana, Slovenia}
\author{Martin Eckstein}
\affiliation{Department of Physics, University of Erlangen-N\"urnberg, 91058 Erlangen, Germany}

\date{\today}

\begin{abstract}
We present a formalism based on nonequilibrium dynamical mean field theory (DMFT) which allows to compute the time-resolved X-ray absorption spectrum (XAS) of photo-excited  solids. By applying this formalism to the photo-doped half-filled and quarter-filled two-orbital Hubbard models in the Mott insulating regime
we clarify how the time-resolved XAS signal reflects the nonequilibrium population of different local states. Apart from the missing broadening associated with continuum excitations, the atomic XAS spectrum computed with the nonthermal state populations provides a good approximation to the full nonequilibrium DMFT result. This suggest a route to combine the accurate DMFT description of nonequilibrum states of solids with cluster calculations of the XAS signal.     
\end{abstract}

\maketitle

\section{Introduction}

X-ray absorption spectroscopy (XAS) allows to explore the local electronic configuration in solids. It is element-specific and has a sufficiently high energy resolution to 
distinguish different valence states 
of a given atom. Together with the charge conserving nature of the probe, this makes XAS an attractive technique for the study of Mott insulators and correlated materials.\cite{derLann1986,Himpsel1991,Haverkort2004}  
XAS is particularly appealing for the investigation of photo-excited nonequilibrium states,\cite{Basov2017, Claudio2016,Torre2021} since new generation free-electron laser sources enable time-resolved XAS (trXAS) measurements with high energy resolution,\cite{Baykusheva2022} and provide information on the nonequilibrium electronic structure which is  complementary to time-resolved photoemission. 
Since the X-ray absorption process represents the first step in resonant inelastic X-ray scattering (RIXS) experiments, a better understanding of trXAS is also relevant in the context of time-resolved RIXS, which can reveal the nonequilibrium physics of low-energy excitations.\cite{Mitrano2020}

XAS on correlated materials can often be well understood in a local picture, because 
its final state typically involves 
strongly bound excitons formed by the localized core hole and the valence electrons. The spectra can 
therefore 
be described to a good approximation using small clusters obtained within ligand field theory,\cite{Haverkort2012} which 
gives 
access to local properties such as  
multiplet energies, crystal-field excitations, etc. In a photo-excited system out of equilibrium, we expect that 
fundamental dynamical processes will be reflected in these local properties. For example, the redistribution of charges between different orbitals will modify the screening of local interactions (Hubbard $U$) and change the electrostatic interaction between the orbitals (Hartree band shifts). This can give rise to level shifts\cite{Wegkamp2014, Golez2019, TancogneDejean2018} which should be directly reflected in the XAS signal.\cite{Baykusheva2022} In multi-orbital systems we furthermore expect an ultrafast change of local multiplet occupations. For example, in a degenerate two-orbital Mott insulator at half-filling, the dominant local multiplet in equilibrium is the high-spin state, but after a laser excitation the photo-doped carriers can assist the ultrafast creation of other multiplets (so called Hund's excitons, i.e., inter- and intra-orbital singlets) which can be relatively long lived.\cite{Strand2017, Rincon2018, Petocchi2019, Gillmeister2020} 

Due to its local nature, trXAS would be ideal to reveal these ultrafast local processes, which are likely to influence also photo-induced phase transitions and hidden phases. Yet, it is not straightforward to adapt the conventional cluster interpretation of XAS to the time-domain. 
Even if one can 
define a finite cluster that captures the dynamics of a local exciton, there is no 
simple principle which defines 
the nonequilibrium initial state of this cluster before the X-ray absorption, in contrast to equilibrium, where the initial state is the cluster ground state. A pragmatic approach would be to use a small cluster (maybe even a single atom) with renormalized parameters and with an adjustable initial quantum state to compute the spectrum and then determine the actual time-dependent configuration by comparison to the experimental spectrum at a given probe time. 
Such an approach would, however, need to be supported by theory. 

In equilibrium, dynamical mean-field theory (DMFT)\cite{Georges1996} provides 
a strategy to construct an optimal embedded cluster model: DMFT   
maps the correlated lattice problem onto an 
Anderson impurity model, 
which can directly be used to compute X-ray spectroscopy signals.\cite{Cornaglia2007} In particular, the combination of 
ab-initio input  from density-functional theory with DMFT 
has enabled 
material-specific calculations for both XAS and RIXS.\cite{Haverkort2014, Lueder2017, Hariki2018, Hariki2020,  Higashi2021} Along the same lines, nonequilibrium DMFT \cite{Aoki2014}  simulations directly provide a suitable nonequilibrium impurity problem from which X-ray spectroscopies can be calculated. This approach has been used recently to compute time-resolved RIXS within nonequilibrium DMFT \cite{Eckstein2021} for  photo-doped Hubbard models and electron-phonon coupled systems.\cite{Werner2021a, Werner2021b} In these simulations, a nonequilibrium impurity-problem with a self-consistently determined DMFT bath is coupled to a core level. The 
core-valence excitation 
by the X-ray pulse is then obtained from time-dependent response functions analogous to 
what would have to be computed in  
exact diagonalization studies, \cite{Chen2019, Wang2020} but for the continuous environment.

In this paper, we discuss how the DMFT approach can be used to compute the trXAS signal, 
focusing on the half-filled and quarter-filled two-band Mott insulator as a representative example. 
The explicit time-dependent numerical calculations support the simpler above-mentioned interpretation based on few-site or atomic clusters in combination with nonequilibrium initial states, i.e.,  the atomic XAS signals computed with the non-thermal populations of the time-evolving state provide a good approximation of the nonequilibrium DMFT result. trXAS therefore allows to track the nonequilibrium populations of the local many-body states. In addition, 
a change of the line broadening out of equilibrium can be related to additional decay channels for the excitonic final states, which are activated by photo-excitation.

The paper is organized as follows: In Sec.~\ref{sec:formalism} we present the general formalism and the observables needed to calculate the time-dependent XAS signal. This section is closely related to the formalism derived in the context of RIXS in Ref.~\onlinecite{Eckstein2021}, but adapted here to the measurement of the XAS signal. Section~\ref{sec:2orbital} presents the applications of the nonequilibrium DMFT-based XAS formalism to the photo-excited half-filled and quarter-filled two-orbital Hubbard model, and the analysis of the time-dependent spectra. Section~\ref{sec:conclusions} contains a summary and conclusions. 

\section{Formalism}
\label{sec:formalism}

\subsection{General remarks}
\label{sec:general}

As in the equilibrium DMFT schemes for calculating X-ray scattering and absorption\cite{Haverkort2014, Lueder2017, Hariki2018, Hariki2020} a core level is added to the DMFT impurity model, and the spectra are computed for a fixed hybridization function. The trXAS simulation thus involves two steps: (i) A nonequilibrium DMFT simulation 
determines the time-dependent valence band hybridization function which represents the pump-induced nonequilibrium state of the lattice system, 
and (ii), a postprocessing step 
simulates the core-valence excitation due to the X-ray probe pulse. 
To limit the core hole life-time, a completely filled electron bath will be added to the core level. 
We start by describing the second step for an atomic system without core bath, and then explain how to add the hybridization function and the core bath in the Keldysh action formulation.  

\subsection{Local Hamiltonian}

The local part of the impurity Hamiltonian is given by
\begin{align}
\label{hloc}
H_\text{loc}=
H_d 
+
H_c
+
H_{cd}
+
H_\text{dip},
\end{align}
where $H_d$ describes the valence orbitals, with annihilation operators $d_{\gamma,\sigma}$ for orbital $\gamma$ and spin $\sigma$. The number of orbitals and the local interaction between the electrons in these valence orbitals is arbitrary within our formalism and need not be specified at this moment. $H_c$ is the Hamiltonian of the core level(s). To simplify the notation, we restrict the discussion to a single orbital with energy $\epsilon_c$, and creation (annihilation) operators $c_{\sigma}^\dagger$ ($c_\sigma$),
\begin{align}
H_{c}
=\epsilon_{c}\sum_{\sigma} c_{\sigma}^\dagger c_{\sigma}.
\end{align}
We describe the core-valence interaction $H_{cd}$ by the density-density interaction
\begin{align}
H_{cd} = U_{cd} \,n_cn_d,
\end{align}
where $n_c = \sum_{\sigma} c_{\sigma}^\dagger c_{\sigma}$ and $n_d=\sum_{\sigma,\gamma} d_{\gamma,\sigma}^\dagger d_{\gamma,\sigma}$ are the total core and valence occupation. The incoming photon mode with energy $\omega_{\text{in}}$ and polarization $\bm \epsilon_{\text{in}}$ is treated semi-classically. The electric field at the impurity site is given by ${\bm E}(t) = g\bm\epsilon_{\text{in}} [s(t)e^{-i\omega_{\text{in}}t}+\text{H.c.}]$, where $s(t)$ is the probe envelope that defines the time window in which the probe is acting on the solid, and 
the amplitude $g$ is introduced to later consider the limit of a weak pulse. 
The Hamiltonian for the dipolar light-matter interaction may thus be written as
\begin{align}
\label{hidip}
H_\text{dip}
=
g \Big(s(t)e^{-i\omega_{\text{in}}t}+\text{H.c.} \Big)
 \Big(\sum_{\gamma,\sigma}
p_{\gamma}^{\text{in}}  
P_{\gamma,\sigma}^\dagger + \text{H.c.} \Big),
\end{align}
where the operator $P_{\gamma,\sigma}^\dagger=d_{\gamma,\sigma}^\dagger c_{\sigma}$ creates a core hole by transferring an electron with spin $\sigma$ to orbital $\gamma$, and $p_{\gamma}^{\text{in}}=\langle d_{\gamma}| \bm\epsilon_{\text{in}}\cdot  \bm r  | c \rangle$ is the dipolar transition matrix element for the mode with polarization $\bm\epsilon_{\text{in}}$. 
For later use, we also introduce the total excitation operator 
\begin{align}
P^\dagger=
\sum_{\gamma\sigma} 
p_{\gamma}^{\text{in}}   P_{\gamma,\sigma}^\dagger.
\end{align}

\subsection{DMFT embedding}

To describe the lattice system, we add to the local model Eq.~\eqref{hloc} the valence hybridization function obtained from a DMFT calculation. Because we are interested in time-dependent processes, we use the Keldysh formalism with a closed time contour $\mathcal{C}$.\cite{Aoki2014} The  impurity model can be described by its action
\begin{align}
\label{simp}
\mathcal{S}_\text{imp} =
\mathcal{S}_\text{loc}
+
\mathcal{S}_{dd}
+
\mathcal{S}_{cc},
\end{align}
where the local 
contribution 
is given by
\begin{align}
\mathcal{S}_\text{loc} = -i \int_{\mathcal{C}} dt  \,
H_\text{loc}(t),
\end{align}
and the second term,
\begin{align}
\label{sdd}
\mathcal{S}_{dd} = -i \int_{\mathcal{C}} dt \,dt' \,\sum_{\sigma,\gamma,\gamma'} d_{\gamma,\sigma}^\dagger (t) \Lambda_{\gamma,\gamma',\sigma}(t,t') d_{\gamma',\sigma}(t'),
\end{align}
with the hybridization function $\Lambda_{\gamma,\gamma',\sigma}(t,t')$, describes the hybridization of the valence orbitals with the self-consistent bath representing the embedding in the lattice. The last term $\mathcal{S}_{cc}$ describes a particle reservoir coupled to the core level, which is explained in Sec.~\ref{sec:core} below.

Because the core-valence interaction is local, the electronic structure of the valence bath is not affected by the XAS process itself. Thus, as mentioned in Sec.~\ref{sec:general}, the hybridization function  
$\Lambda$ 
in the time-dependent formalism is computed in a separate nonequilibrium DMFT simulation without core hole excitation, which includes the pump laser fields or other excitations that drive the system out of equilibrium. The XAS signal is subsequently evaluated from the impurity model \eqref{simp} with the given $\Lambda_{\gamma,\gamma',\sigma}(t,t')$. 

\subsection{Core bath}
\label{sec:core}

The core hole typically has a very short lifetime due to Auger decay. This decay is often taken into account phenomenologically, for example by adding an exponential decay with rate $\Gamma$ to the core hole state,\cite{Chen2019, Hariki2020, Wang2020} which implies a Lorentzian broadening of the spectra. In our real-time formalism, which explicitly simulates the X-ray absorption process, the decay must be included explicitly. We achieve this by coupling a particle reservoir to the core level, which yields an additional term
\begin{align}
\mathcal{S}_{cc} = -i \int_{\mathcal{C}} dt \,dt'\, \sum_{\sigma} c_{\sigma}^\dagger (t) \Delta_c(t,t') c_{\sigma}(t')
\label{bathcc}
\end{align}
with a core hybridization function $\Delta_c(t,t')$ in the action \eqref{simp}. This core hybridization function represents an entirely filled reservoir of electrons, which can lead to the decay of a core hole, but cannot create core holes. For a wide core bath density of states (DOS), this core bath exactly reproduces the conventional exponential decay when our real-time formalism is applied to equilibrium systems (see Ref.~\onlinecite{Eckstein2021} for a more detailed discussion of this point). In Appendix~\ref{app:tests},  we provide the precise form of $\Delta_c$.  
For a core bath with finite bandwidth the functional form of the core hole population decay can be different from exponential, but in this context it should be noted that also the conventional exponential core-hole decay is essentially a heuristic approximation.

\subsection{X-ray absorption}
\label{sect_wo_rwa}

To evaluate the time-resolved X-ray absorption, we start from a general expression for the energy absorption rate, 
\begin{align}
A(t) \equiv \frac{d}{dt} \langle H(t) \rangle_{\!H} = \langle \partial_t H(t)\rangle_{\!H}.
\end{align}
Here $H$ is the full Hamiltonian, which includes the local part \eqref{hloc}, and the environments which were expressed in an action form in Eqs.~\eqref{sdd} and \eqref{bathcc}.  
The X-ray absorption is obtained from the terms in the partial time derivative $\partial_t H(t)$ which are related to the dipolar interaction,
\begin{align}
\label{xasrate}
A_{\text{XAS}}(t) = 
g\big\langle P^\dagger(t) + P(t)\big\rangle_{\!H}
\frac{\partial}{\partial t}\Big(s(t)e^{-i\omega_{\text{in}}t}+\text{H.c.}\Big).
\end{align}
The XAS signal is given by the absorbed photon number, i.e., the total absorbed X-ray energy per photon energy, to leading order in $g$,
\begin{align}
\label{ixas1}
I_{\text{\rm XAS}} = 
\lim_{g\to 0} \frac{1}{g^2\omega_{\text{in}}}\int_{-\infty}^\infty dt \,A_{\text{XAS}}(t).
\end{align}
This expression can be directly evaluated using the numerical techniques which are employed for the solution of the impurity model. For the applications in this work we use the real-time non-crossing approximation to the hybridization expansion\cite{Eckstein2010} and the NESSi simulation package to deal with Keldysh nonequilibrium Green's functions.\cite{Schueler2020} 

Usually, there is a large energy  scale separation between the bandwidth of the valence band, which is at most of the order of a few eV, and the energies $\omega_{\text{\text{in}}}$, and $|\epsilon_{c}|$, which can be of the order of  keV. In this case, the numerics can be simplified by a rotating wave approximation, which would remove the need to simulate the fast oscillating pulse. The corresponding equations are summarized in Appendix~\ref{app:rotwave}.

\section{Two-orbital Hubbard model}
\label{sec:2orbital}

\subsection{Setup}

In this section, we compute time-resolved XAS signals for a two-orbital Hubbard model, using the nonequilibrium DMFT based formalism outlined in Sec.~\ref{sec:formalism}.  The local Hamiltonian of the two-orbital model reads
\begin{align}
H_\text{loc}=&\sum_{\alpha} U n_{\alpha\uparrow}n_{\alpha\downarrow} \nonumber\\
&+ \sum_{\sigma}[(U-2J) n_{1\sigma}n_{2\bar\sigma}+(U-3J)n_{1\sigma}n_{2\sigma}]\nonumber\\
&+ U_{cd} n_cn_d+\epsilon_c n_c-\mu(n_d+n_c),
\label{ham}
\end{align}
where $\alpha=1,2$ 
labels 
the valence orbitals and $\sigma=\uparrow,\downarrow$ denotes spin. The annihilation and density operators for the valence states are  $d_{\alpha\sigma}$ and $n_{\alpha\sigma}$, while the annihilation and density operators for the core are denoted by $c_\sigma$ and $n_{c\sigma}$. The total densities are $n_d=\sum_{\alpha,\sigma}n_{\alpha\sigma}$ and $n_c=\sum_\sigma n_{c\sigma}$, respectively, and the parameter $\epsilon_c$ allows to shift the energy of the core level relative to the valence orbitals. $U$ is the intra-orbital interaction, $J$ the Hund coupling, $U_{cd}$ the interaction between core and valence electrons, and $\mu$ the chemical potential. 

For the analysis of the spectra, the local multiplet states, which are listed in Table~\ref{tabstates01}, will be relevant. In the doubly occupied sector, we denote the high-spin states by $2_h$, the low spin states with two electrons in different orbitals by $2_l$ and the intra-orbital singlet by $2_s$. 
The other multiplets will be denoted by $m_d$, where $m\in\{0,1,3,4\}$ is the $d$-occupation. Furthermore, we will denote by an underbar ($\underbar 0_d$, $\underbar 1_d$, $\underbar 2_h$, ...) a local state with the given  $d$-orbital  configuration and a core hole, and by $E(\alpha)$ and $E(\underline{\alpha})$ the respective energies, which can be directly obtained from Eq.~\eqref{ham}.

\begin{table}
\begin{tabular}{l|l|l}
\hline
$\alpha$ & $|j\rangle\in\alpha$ & $E_\text{int}(\alpha)$\\
\hline\hline
$0_d$ \hspace{5mm} & $|0,0\rangle$ & 0\\
\hline
$1_d$ & $|\!\uparrow,0\rangle$, $|\!\downarrow,0\rangle$, $|0,\uparrow\rangle$, $|0,\downarrow\rangle$ & 0\\
\hline
$2_h$ & $|\!\uparrow,\uparrow\rangle$, $|\!\downarrow,\downarrow\rangle$ & $U-3J$\\
\hline
$2_l$ & $|\!\uparrow\downarrow,0\rangle$, $|0,\uparrow\downarrow\rangle$ & $U-2J$\\
\hline
$2_s$ & $|\!\uparrow,\downarrow\rangle$, $|\!\downarrow,\uparrow\rangle$ &$U$\\
\hline
$3_d$ & $|\!\uparrow,\uparrow\downarrow\rangle$, $|\!\downarrow,\downarrow\uparrow\rangle$,$|\!\uparrow\downarrow,\uparrow\rangle$, $|\!\downarrow\uparrow,\downarrow\rangle$ \hspace{5mm} & $3U-5J$\\
\hline
$4_d$ & $|\!\uparrow\downarrow,\uparrow\downarrow\rangle$& $6U-10J$\\
\end{tabular}
\caption{Labeling of the local $d$-multiplet states. The second column shows the $d$ states belonging to a given multiplet $\alpha$, where the first (second) entry in the ket refers to the occupation of orbital $1$ ($2$). The last column shows the interaction energy.  The total energy is $E(\alpha) = E_\text{int}(\alpha)+E(n_d,2)$ with the density contribution $E(n_d,n_c) = U_{cd} n_cn_d+\epsilon_cn_c-\mu(n_d+n_c)$. The energy for the corresponding core-excited state $\underline{\alpha}$ is  $E(\underline{\alpha}) = E_\text{int}(\alpha)+E(n_d,1)$.}
\label{tabstates01}
\end{table}   

The kinetic 
energy 
of the valence electrons is $H_\text{kin}=v(t)\sum_{\langle i,j\rangle,\gamma,\sigma} (d^\dagger_{i\gamma\sigma}d^{\phantom\dagger}_{j\gamma\sigma}+ \text{H.c.})$, which describes hopping with amplitude $v(t)$ between nearest neighbors on a lattice; spin $\sigma$ and orbital $\gamma\in 1,2$ are preserved in this hopping process. The DMFT simulations are implemented for an infinite-dimensional Bethe lattice with appropriately renormalized hopping amplitude. In this case, the DMFT self-consistency relation directly connects the impurity hybridization function $\Lambda_{\gamma,\gamma'}$ in the action \eqref{simp} and the local Green's functions $G_{\gamma,\gamma'}$ with the closed-form self-consistency relation \cite{Georges1996} $\Lambda_{\gamma,\gamma'}(t,t') = \delta_{\gamma,\gamma'} v(t) G_{\gamma,\gamma}(t,t') v(t')^*$. 
There is no dependence on spin $\sigma$, and $\Lambda$ is orbital-diagonal because the hopping is diagonal. 
The X-ray pulse will lead to off-diagonal components  in the Green's function, which measure the current between the core and valence orbitals, but this does not affect $\Lambda$, since the hybridization function is fixed by the system which is not perturbed by the X-ray pulse.

To ``photo-excite" the system, we employ a time-dependent modulation of the hopping parameter $v$,
\begin{equation}
v(t)=v_0+v_{\rm pu}s_{\rm pu}(t-t_{\rm pu})\sin[\omega_{\rm pu}(t-t_{\rm pu})]. 
\end{equation}
Here, $\omega_{\rm pu}$ is the frequency of the pump pulse, $v_{\rm pu}$ its maximum amplitude and $s_{\rm pu}(t-t_{\rm pu})$ the envelope function centered at time $t_{\rm pu}$. In the following, we will set $v_0=1$ as the unit of energy ($\hbar/v_0$ as the unit of time), which corresponds to a noninteracting lattice DOS of bandwidth $4$. The initial inverse temperature is $\beta=5$.

To compute the XAS signal, we follow  Sec.~\ref{sect_wo_rwa} and solve the 3-orbital impurity model given by the valence-band 
Anderson impurity model 
supplemented by the core level [c.f.~Eq.~\eqref{ham} for the local part]. The probe pulse is described by a dipolar coupling between the core level and the $d$-orbital with index $\gamma=1$: $H_{\rm pr}=E_{\rm pr}(t)(P+P^\dagger)$, with $P^\dagger=\sum_\sigma d^\dagger_{1\sigma}c^{\phantom\dagger}_\sigma$. (Due to the degeneracy between the valence orbitals in our model, the dipolar coupling to the orbital $\gamma=2$ would give the same results.)
The probe field has the form 
\begin{equation}
\label{probeulse}
E_{\rm pr}(t)=g\,s_{\rm pr}(t-t_{\rm pr})\sin[\omega_\text{in}(t-t_{\rm pr})],
\end{equation} 
with the probe pulse frequency $\omega_\text{in}$, and an envelope function $s_{\rm pr}(t)=\exp(-t^2/\sigma^2)$  centered at time $t_{\rm pr}$.  
Unless otherwise stated, we use probe pulses with $\sigma^2=8$.   The amplitude $g$ $(=0.01)$, which incorporates the dipole matrix element, is chosen weak enough that the absorbed energy is quadratic in $g$ [c.f.~Eq.~\eqref{ixas1}]. The results below are based on an explicit treatment of this model with a relatively shallow core level, which is however still well separated in energy from the valence states.

The explicit form of the core bath (a box shaped density of states of width $W=20$) is given in Appendix~\ref{app:tests}. As long as $U_{cd}\lesssim W/2$, the core-hole decay induced by this bath will lead to an additional broadening of the XAS signal, on top of the broadening resulting from the short probe pulse. 
The corresponding lineshape of an isolated transition is analyzed in Fig.~\ref{fig_atomic} of Appendix~\ref{app:tests}. The power spectrum $|s_{\rm pr}(\omega)|^2 \sim e^{-\omega^2 \sigma^2/2}$ of the probe pulse is a 
Gaussian. The lineshape for the single XAS absorption line then results from a convolution of the core-hole decay and this finite energy resolution of the probe. To a good approximation, the resulting lineshape is again Gaussian 
\begin{align}
\label{atomicline}
S_\text{at}(\omega) \sim e^{-\omega^2 \sigma_{\rm eff}^2/2},
\end{align}
with $\sigma_{\rm eff}^2=5.5$ for the probe pulses with $\sigma^2=8$
and $\sigma_{\rm eff}^2=1.6$ for the shorter probe pulses with $\sigma^2=2$ (see Appendix~\ref{app:tests}). 
For simplicity we will use this Gaussian lineshape in the analysis below, unless otherwise indicated.

\subsection{Half-filled model} 
\label{sec_half}
 
\begin{figure}[t]
\begin{center}
\includegraphics[angle=-90, width=0.45\textwidth]{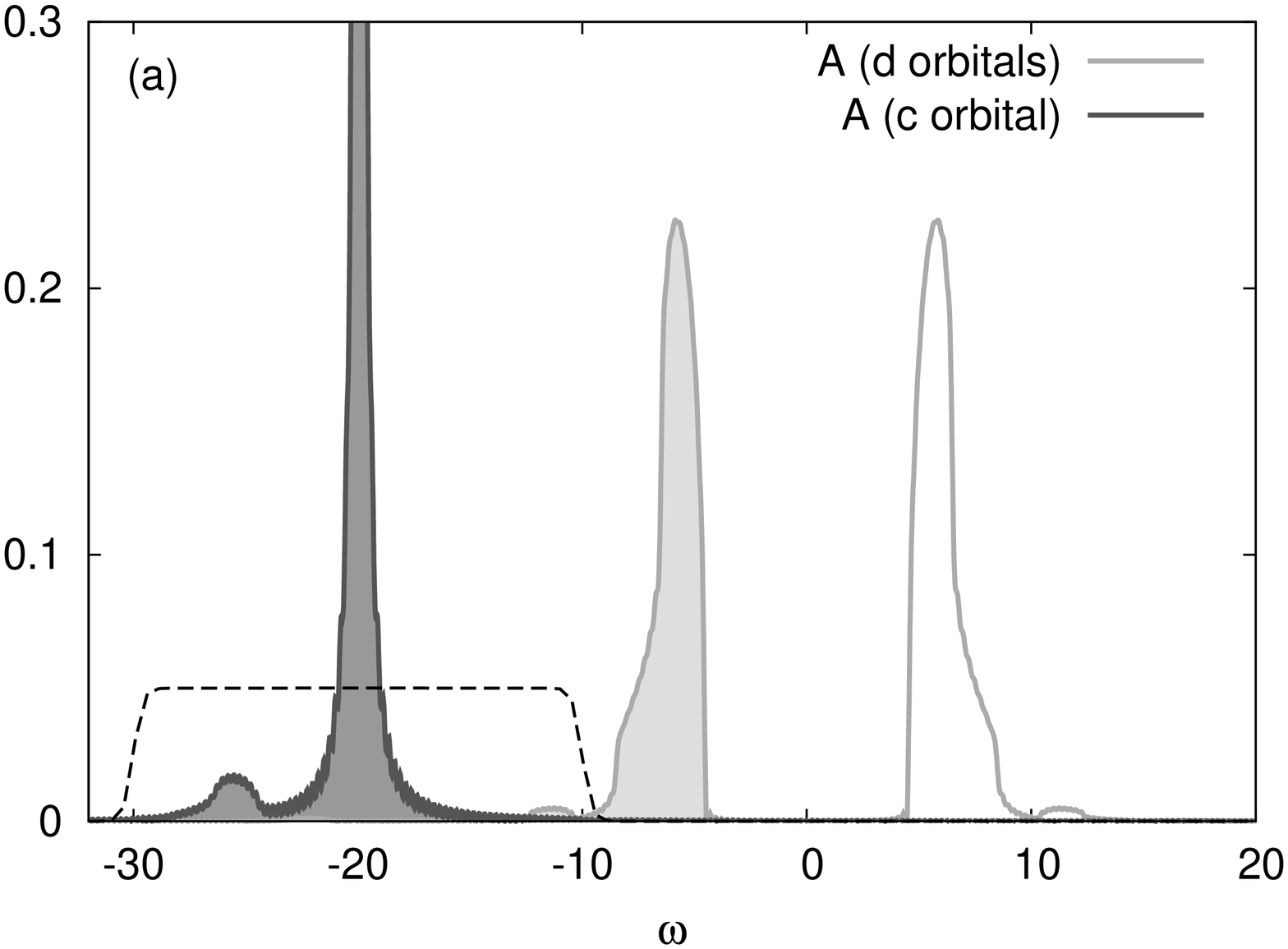}\\
\hspace{0.7mm}
\includegraphics[angle=-90, width=0.442\textwidth]{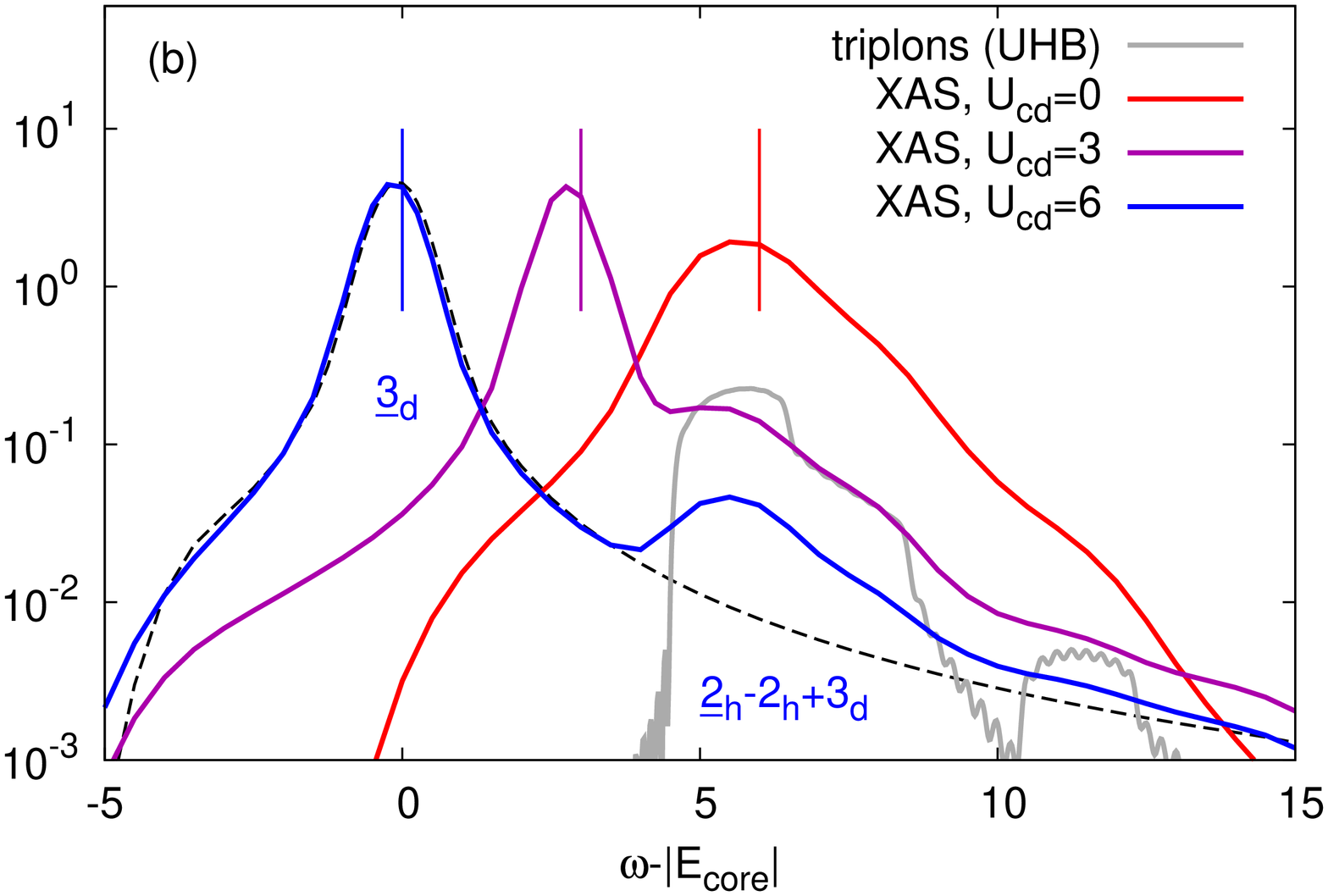}
\caption{
(a) Equilibrium spectral functions $A(\omega)$ of the half-filled model with $U=10$, $J=2$, $U_{cd}=3$ and $E_\text{core}=-20$. The coupling to a Fermion bath with box-shaped DOS with bandwidth $W=20$ (dashed black line) and coupling strength $v_\text{bath}=1$ leads to a broadening of the $c$-DOS. The main Hubbard bands in the $d$-DOS are near $\omega=\pm 6$, with satellites at $\pm 8$ and  $\pm 12$. (b) XAS signal as a function of $\omega-|E_\text{core}|$, for indicated values of $U_{cd}$. Here, we use the long probe pulse with $\sigma^2=8$. 
Vertical lines  correspond to the bare exciton energies $E_\text{ex}=E(\underline{3}_d)-E(2_h)$, the gray line shows the shifted upper Hubbard band, which  indicated the energy range for the lowest continuum excitation $\underline{2}_h-2_h+3_d$, and the black dashed line shows the lineshape $S_\text{at}(\omega-E_\text{ex})$ of the atomic transition $2_h\rightarrow \underline 3_d$.
}
\label{fig_half}
\end{center}
\end{figure}

\subsubsection{Spectral functions}
 
We first discuss the half-filled two-orbital Hubbard model and choose the interaction parameters of the $d$ manifold as $U=10$, $J=2$.  The equilibrium spectral functions $A_\alpha(\omega)=-\frac{1}{\pi}\text{Im}G^R_\alpha(\omega)$ are shown for $\alpha=c,d$ in Fig.~\ref{fig_half}(a). Because of the large Hund coupling, the half-filled high-spin states $2_h$ dominate the equilibrium $d$ configuration. The upper and lower Hubbard band is centered around the electron addition energy  $E_\text{UHB}=E(3_d)-E(2_h)$ and electron removal energy  $E_\text{LHB}=-[E(1_d)-E(2_h)]$, respectively,  
so that 
$E_\text{UHB}-E_\text{LHB}=U+J$. In the following, we will fix $\mu$ such that the $d$-electron addition and removal energies for the state $2_h$ are symmetric ($E_\text{UHB}=-E_\text{LHB}$), which implies $\mu=2U_{cd} + (3U-5J)/2$. The core band is centered around the core removal energy $-[E(\underbar 2_h)-E(2_h)]\equiv E_\text{core}$; 
 we will fix $\epsilon_c$ such that $E_\text{core}=-20$, which implies $\epsilon_c=\mu-2U_{cd}+E_\text{core}$. In a realistic system, the energy splitting to the core level is much larger, but for the calculation of the XAS spectrum with the present formalism, we merely need a completely occupied core level that is sufficiently separated from the $d$ levels. All the XAS results will be plotted as a function of $\omega_\text{in}-|E_\text{core}|$.

The width of the upper (lower) Hubbard bands basically gives the range of kinetic energy for the triplons $3_d$ (singlons $1_d$), which are delocalized states if the core is filled ($n_c=2$). In addition to the main Hubbard bands at $\omega\approx \pm(U+J)/2=\pm 6$ we recognize shoulder and satellite structures at higher energies ($\omega\approx \pm 8$ and $\pm 12$), which are more clearly seen on the log-scale plot in Fig.~\ref{fig_half}(b) (gray line). These structures originate from local spin excitations on neighboring sites (triplon insertion plus hopping to a neighboring site, which leaves behind a low-spin doublon \cite{Lysne2020}), and would not be present in the atomic limit.

\subsubsection{XAS in equilibrium}

We now turn to the XAS signal, shown in Fig.~\ref{fig_half}(b). As in the case of the spectral function, the local multiplet energies are a good starting point for analyzing the main XAS features.  Starting from the dominant $2_h$ state, one can create a local configuration $\underbar 3_d$ by transferring an electron from the core to the $d$ shell. In the atomic limit, there would be a peak in the XAS signal at $\omega= E(\underbar 3_d)-E(2_h) = (U+J)/2-U_{cd}-E_\text{core}\equiv E_\text{ex}$, broadened by the core lineshape and the probe envelope. Due to virtual charge fluctuations, the true initial state before the X-ray absorption has admixtures from $1_d$, with a triplon $3_d$ on a different site. This makes final states accessible with a configuration $\underbar 2_x-2_h+3_d$, where $x\in{h,l,s}$, and the addition ``$-2_h+3_d$'' indicates that in addition to the local state a delocalized triplon $3_d$ has been created from a $2_h$ state. In the following, we will call the final state $\underbar 3_d$  the {\em exciton}, while the final states which involve a delocalized excitation will be referred to as the {\it continuum}. The lowest continuum band, $\underbar 2_h-2_h+3_d$, is at $E_\text{cont}= E(\underbar 2_h)-E(2_h) +\epsilon_t = \epsilon_t -E_\text{core}$, where $\epsilon_t$ is the energy of the delocalized triplon, which lies in the upper Hubbard band region. 
The exciton is therefore simply shifted by a binding energy $E_{\text{bind}}=-U_{cd}$ with respect to the center of the continuum at $(U+J)/2-E_\text{core}$.

Figure \ref{fig_half}(b) shows the XAS signal for three different values of $U_{cd}$, together with the bare exciton energies $E_\text{ex}$ (vertical lines) and the energy range of the continuum $\underline{2}_h-2_h+3_d$ (shifted upper Hubbard band; gray line). As expected, for large $U_{cd}$, when the exciton does no longer overlap with the continuum, the exciton becomes an increasingly sharp feature and the continuum is strongly suppressed, while for small $U_{cd}$ the two signals merge into a broad feature. In the strongly localized regime of large $U_{cd}$, the shape of the exciton approaches the lineshape of an isolated transition line $S_\text{at}(\omega)$ (dashed black line), which is determined by the broadening due to the core-hole decay and the limited energy resolution of the short probe. (Here we do not use the Gaussian fit Eq.~\eqref{atomicline} to the line, but the exact atomic lineshape as computed in Appendix~\ref{app:tests}.)

\begin{figure}[tbp]
\begin{center}
\includegraphics[angle=-90, width=0.45\textwidth]{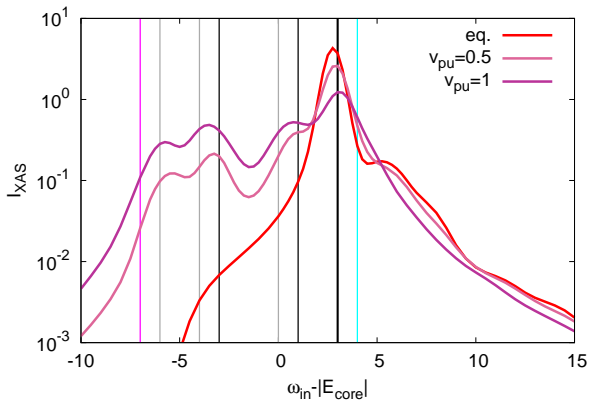}
\caption{
XAS spectra of the equilibrium (eq.) and photo-doped half-filled model with $U=10$, $J=2$, $U_{cd}=3$ on a logarithmic scale. We show nonequilibrium results for $v_{\rm pu}=0.5$ and  $1$, with pump pulse frequency $\omega_{\rm pu}=12$ (resonant excitation). The probe time is $t_{\rm pr}=6.5$ and the probe envelope has $\sigma^2=8$ (see Fig.~\ref{fig_xas_half_states}(a) for the pulse profiles). The vertical lines indicate the energies associated with the following atomic transitions: $2_h\rightarrow \underline{3}_d$ (thick black), $2_{l,s}\rightarrow \underline{3}_d$ (thin black), $1_d\rightarrow \underline{2}_x$ (gray), $3_d\rightarrow \underline{4}_d$ (light blue), $0_d\rightarrow \underline{1}_d$ (pink).
}
\label{fig_xas_half}
\end{center}
\end{figure}

The slight shifts of the exciton peaks with respect to the bare excitation line $E_\text{ex}$ can be interpreted as hybridization shifts of the energies, which would be corrected if the local states would be computed in a cluster with surrounding orbitals rather than for an isolated site. (The DMFT impurity model of course includes this environment.) Moreover, we remark that virtual admixtures to the initial state also allow different final states: There should be a continuum $\underbar 4_d-2h+1_d$, corresponding to a local state $\underbar 4_d$ with one 
additional 
electron and a delocalized hole $1_d$. This signal lies outside the spectral region shown here and is strongly suppressed. Moreover, the $\underbar 3_d$ exciton could be accompanied by satellites corresponding to additional doublon multiplets (Hund's excitations)  on other sites. However, the creation of such satellites involves high-order hopping processes and they are hence also strongly suppressed.

\begin{figure}[t]
\begin{center}
\includegraphics[angle=-90, width=0.45\textwidth]{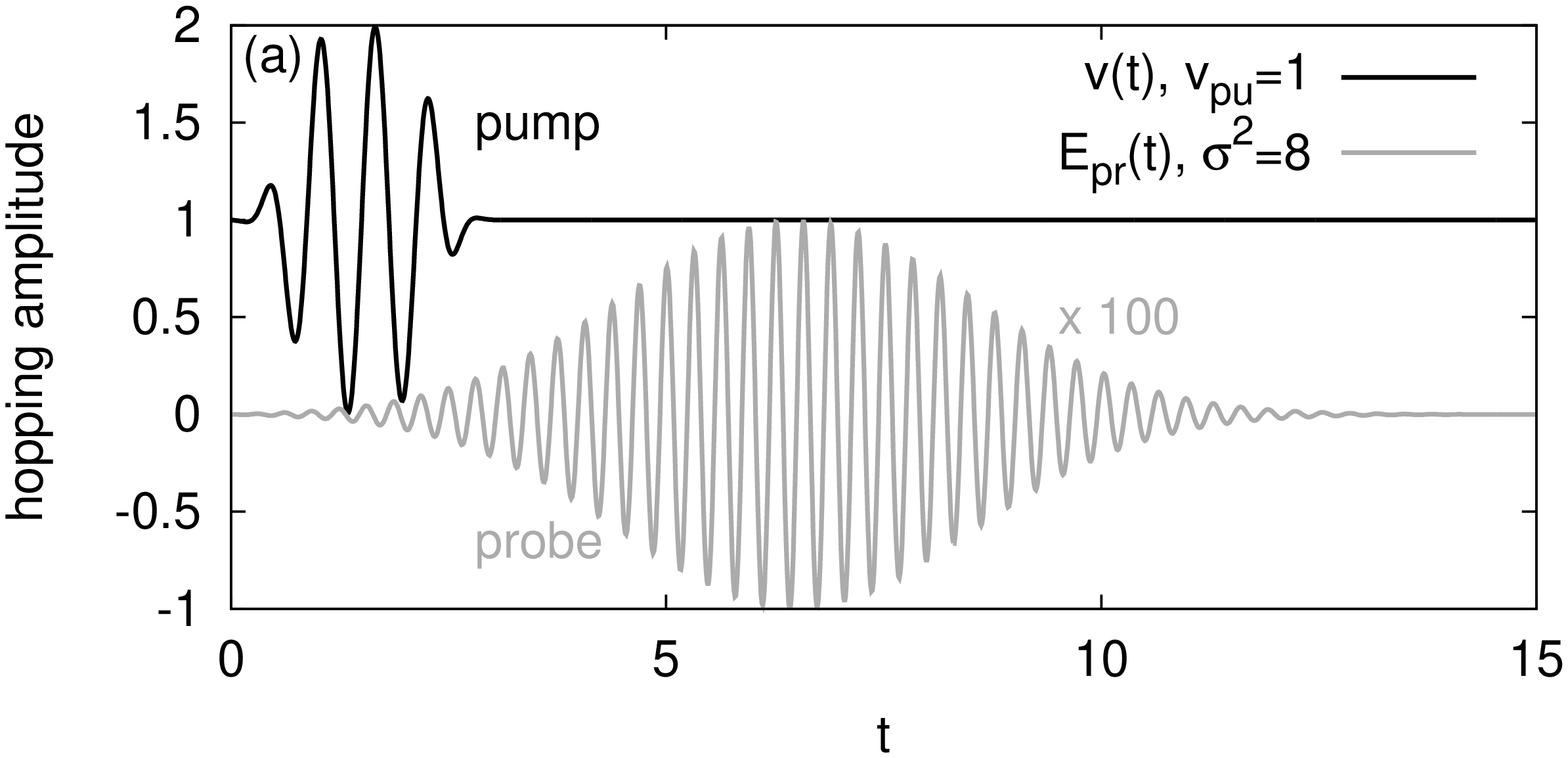}
\includegraphics[angle=-90, width=0.45\textwidth]{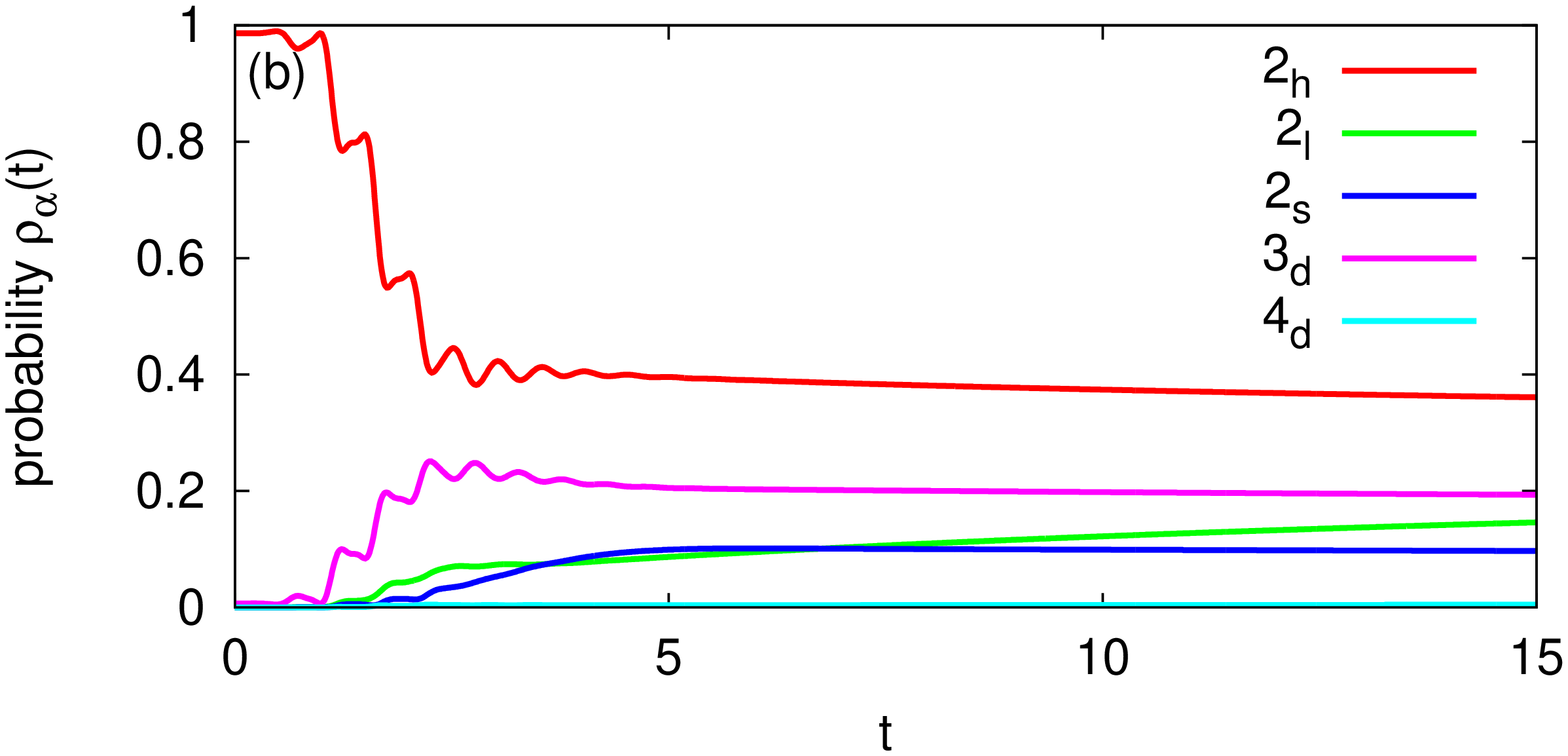} 
\caption{
(a) Illustration of the pump pulse (amplitude $v_\text{pu}=1$, $t_\text{pu}=1.5$, $\omega_\text{pu}=12$) and of the longer probe pulse with $\sigma^2=8$ (amplitude $g=0.01$, $t_\text{pr}=6.5$, $\omega_\text{in}=20$).
(b) Probability of the different local states in the half-filled model with $U=10$, $J=2$, $U_{cd}=3$, which is excited by a pump pulse with $v_{\rm pu}=1$ and $\omega_{\rm pu}=12$.  
Because of particle-hole symmetry, triplon and singlons as well as empty sites and quadruplons have equal occupation ($\rho_{1_d}=\rho_{3_d}$ and $\rho_{0_d}=\rho_{4_d}$).
}
\label{fig_xas_half_states}
\end{center}
\end{figure}

\subsubsection{XAS after photodoping}

We next discuss the XAS signal of the pump-excited system (resonant excitation with $\omega_{\rm pu}=U+J=12$, probe pulse centered at $t_\text{pr}=6.5$; see Fig.~\ref{fig_xas_half_states}(a) for the pump and probe pulses). As shown in Fig.~\ref{fig_xas_half}, the main peak of the equilibrium system is weakened, while additional spectral weight appears mainly at lower energies, but to some extent also at higher energies. To interpret this rather rich nonequilibrium spectrum, we can again start from a local picture. 
 
\begin{table*}
\begin{tabular}{|l||lr||c||lr|lr|lr|lr|lr|}
\hline
$\alpha_i$ & $\underline {\alpha}$, & $E_{ex}$ & $M_{\alpha_i;\underline{\alpha}}$ & $\underline{\alpha}' - \beta_i + \beta$, & $\Delta E$ &&&&&&&& \\
\hline
\hline
$0_d$ & $\underbar 1_d$, & -7&  $2$  & $\underbar 2_h$$-$$1_d$$+$$0_d$, & -1&$\underbar 2_l$$-$$2_h$$+$$1_d$, & 1&$\underbar 2_s$$-$$2_l$$+$$1_d$, & -1&&&&
 \\
\hline
$^{*}1_d$ & $\underbar 2_h$, & -6&  $\frac{1}{2}$  &  $\underbar 1_d$$-$$0_d$$+$$1_d$, & 1&$\underbar 3_d$$-$$3_d$$+$$2_l$, & 1&&&&&&
 \\
\hline
$^{*}1_d$ & $\underbar 2_l$, & -4&  $\frac{1}{2}$  &  $\underbar 1_d$$-$$1_d$$+$$2_h$, & -1&$\underbar 3_d$$-$$2_s$$+$$1_d$, & -1&$\underbar 3_d$$-$$3_d$$+$$2_s$, & -1&&&&
 \\
\hline
$^{*}1_d$ & $\underbar 2_s$, & 0& $\frac{1}{2}$  &  $\underbar 1_d$$-$$1_d$$+$$2_l$, & 1&$\underbar 3_d$$-$$2_l$$+$$1_d$, & -1&&&&&&
 \\
\hline
$2_h$ & $\underbar 3_d$, & 3& $1$  &  $\underbar 2_s$$-$$1_d$$+$$2_l$, & 1&$\underbar 2_l$$-$$1_d$$+$$2_s$, & 1&$\underbar 2_h$$-$$2_l$$+$$3_d$, & -1&$\underbar 2_l$$-$$2_s$$+$$3_d$, & 1&$\underbar 4_d$$-$$3_d$$+$$2_h$, & -1
 \\
\hline
$2_l$ & $\underbar 3_d$, & 1& $1$  &  $\underbar 2_s$$-$$1_d$$+$$2_l$, & 1&$\underbar 2_l$$-$$1_d$$+$$2_s$, & 1&$\underbar 2_h$$-$$2_l$$+$$3_d$, & -1&$\underbar 2_l$$-$$2_s$$+$$3_d$, & 1&$\underbar 4_d$$-$$3_d$$+$$2_h$, & -1
 \\
\hline
$2_s$ & $\underbar 3_d$, & -3& $1$  &  $\underbar 2_s$$-$$1_d$$+$$2_l$, & 1&$\underbar 2_l$$-$$1_d$$+$$2_s$, & 1&$\underbar 2_h$$-$$2_l$$+$$3_d$, & -1&$\underbar 2_l$$-$$2_s$$+$$3_d$, & 1&$\underbar 4_d$$-$$3_d$$+$$2_h$, & -1
 \\
\hline
$^{*}3_d$ & $\underbar 4_d$, & 4& $\frac{1}{2}$  &  $\underbar 3_d$$-$$2_h$$+$$3_d$, & 1&&&&&&&&
 \\
\hline
\end{tabular}                       
\caption{Possible XAS excitations for the half-filled model with $U=10$, $J=2$, $U_{cd}=3$, starting from a given local initial multiplet $\alpha_i$ (left column). The second column shows possible exciton excitations $\underline{\alpha}$, with the corresponding excitation energy $E_\text{ex}=E(\alpha_i)-E(\underline{\alpha})$ (relative to $|E_\text{core}|$). The third column is the matrix element \eqref{matrixelements}. The subsequent columns list continuum states $\underline{\alpha}' -\beta_i + \beta$ which can be obtained by adding or removing a particle from the exciton $\underline{\alpha}$, while making the transition $\beta_i\to\beta$ in the continuum. The second number in the column is the difference $\Delta E$ of the continuum energy $E_\text{cont}=E(\underline{\alpha}')+E(\beta)-E(\beta_i)$ and the exciton $E_\text{ex}$. Only continua with $|\Delta E| < 3$ are shown. States marked by an asterisk experience broadening  because the initial state is delocalized.}
\label{tabstates}
\end{table*}     

The nonequilibrium excitation is generating multiple possible initial configurations in addition to the $2_h$ state: Figure~\ref{fig_xas_half_states}(b) shows the evolution of the local state occupations for a strong pump $v_{\rm pu}=1$. Plotted is the quantity
\begin{align}
\label{rhoprob}
\rho_{\alpha}(t)= \sum_{i\in\alpha} \langle i | \rho_\text{loc}(t) | i\rangle,
\end{align}
which corresponds to the total occupation of a multiplet $\alpha\in \{0_d,1_d,2_h,2_l,2_s,3_d,4_d\}$, and is measured from the reduced density matrix $\rho_\text{loc}$ of the DMFT impurity site. (By construction, $\sum_{\alpha} \rho_\alpha=1$.) The red line shows the $2_h$ occupation, which dominates in the initial equilibrium state. A substantial fraction of these high-spin doublons is converted into other states by the pulse. In particular, the pulse creates a significant density of triplons $3_d$ (and the same density of singlons $1_d$), see pink line. 
A part of the initial kinetic energy of these delocalized excitations 
is then further converted into different multiplet excitations.\cite{Strand2017} Consequently, one can see an increase of the two types of low-spin doublon states, $2_l$ and $2_s$ (green and dark blue lines). Their relative density depends on the pulse amplitude. A significant population of $4_d$ and $0_d$ configurations (identical value, light-blue line) is only produced by a strong pulse. 

We can now ask to what extent this local state dynamics is reflected in the XAS spectrum. 
In a similar way as discussed for the $2_h$ initial states in equilibrium, 
all other  initial configuration $\alpha$ can be transformed by the X-ray pulse either into exciton configurations $\underline \alpha'$, or into various excitation continua and satellites which involve multiplet excitations on other sites.  
 Let us first discuss the signature of the excitons. Table~\ref{tabstates} gives an overview over the possible initial states $\alpha_i$ (first column) and the accessible exciton states $\underline{\alpha}$ (second column) and their energies $E_\text{ex}=E(\underline{\alpha})-E(\alpha_i)$.  
In a first approach to interpret  the trXAS results in Fig.~\ref{fig_xas_half}, 
 one may compute the spectrum for an isolated site, but allowing for all possible initial configurations, as given by the time-dependent probabilities \eqref{rhoprob}. This amounts to the ansatz 
\begin{align}
I_{\rm XAS} (\omega) 
&=\!
\sum_{i,j} \rho_i(t) |\langle j | P^\dagger | i \rangle|^2 \, S_\text{at} \big(\omega-(E_j-E_i)\big),
\end{align}
where $i$ and $j$ are all possible atomic  initial and final states, $\rho_i$ denotes the initial state occupations, $S_\text{at}$ the atomic lineshape approximated by Eq.~\eqref{atomicline}, and $|\langle j | P^\dagger | i \rangle|^2$ the matrix element. Since the energies $E_i$ and $E_j$ depend only on the multiplet, the sum over all states can be rewritten in terms of a sum over initial and final multiplets, 
\begin{align}
I_{\rm XAS} (\omega)&= \!\sum_{\alpha_i,\alpha} \rho_{\alpha_i}(t) M_{\alpha_i; \underline{\alpha}} \, S_\text{at}\big(\omega-E(\underline{\alpha})+E(\alpha_i)\big),
\label{xasreconstruction}
\end{align}
with the accumulated matrix element 
\begin{align}
\label{matrixelements}
M_{\alpha_i; \underline{\alpha}}=\frac{1}{N_{\alpha_i}}\sum_\sigma\sum_{i\in\alpha_i, j\in\alpha} |\langle j |d_{1\sigma}^\dagger | i\rangle |^2.
\end{align}
for all possible transitions between the $\alpha_i$ multiplet and the $\underline \alpha$ exciton.
This matrix element  takes into account that in the present simulation the X-ray pulse couples only to orbital $1$ and is not spin sensitive. The normalization by the number $N_{\alpha_i}$ of states in the initial multiplet is introduced because $\rho_{\alpha_i}$ measures the total occupation of the multiplet, not the occupation of the individual states [c.f.~Eq.~\eqref{rhoprob}]. The matrix elements are listed in the third column of Table~\ref{tabstates}.

\begin{figure}[tbp]
\begin{center}
\includegraphics[angle=0, width=0.5\textwidth]{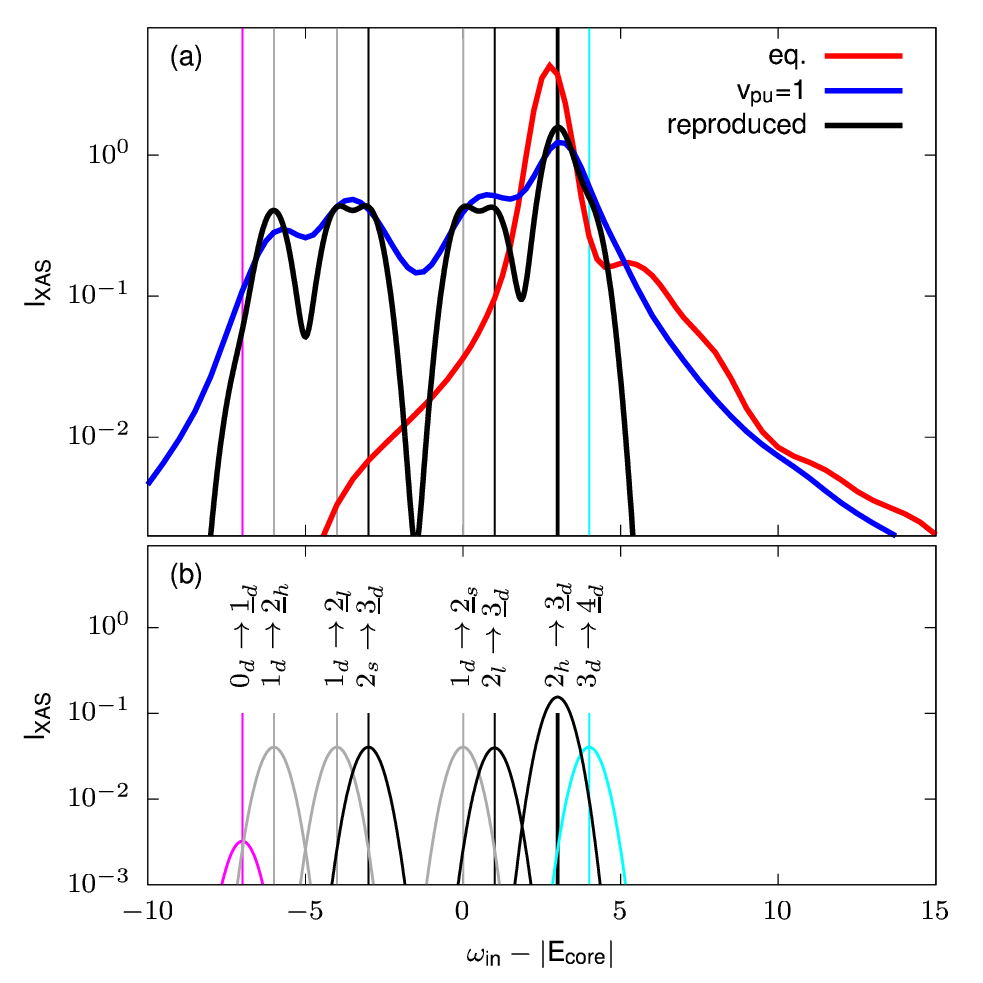} 
\caption{
(a) XAS spectra of the equilibrium and photo-doped half-filled model with $U=10$, $J=2$, $U_{cd}=3$, $v_{\rm pu}=1$ (resonant excitation), measured at $t_{\rm pr}=6.5$ with the long pulse ($\sigma^2=8$). The full result (blue line) is compared to the XAS spectrum reconstructed from the atomic system with the nonequilibrium occupations measured at $t=6.5$ (black line).  The vertical lines indicate the energies associated with the following atomic transitions: $2_h\rightarrow \underline{3}_d$ (thick black), $2_{l,s}\rightarrow \underline{3}_d$ (thin black), $1_d\rightarrow \underline{2}_x$ (gray), $3_d\rightarrow \underline{4}_d$ (light blue), $0_d\rightarrow \underline{1}_d$ (pink), as indicated in panel (b), which also shows the contributions of the different excitons to the spectrum. For the reconstruction, we use 
the lineshape in Eq.~\eqref{atomicline} with 
 $\sigma_\text{eff}^2=5.5$, as measured in the atomic limit.
}
\label{fig_xas_reproduced}
\end{center}
\end{figure}   

In Fig.~\ref{fig_xas_reproduced} we compare the nonequilibrium XAS spectrum for the stronger photoexcitation $v_{\rm pu}=1$, measured at $t_{\rm pr}=6.5$ with the longer probe pulse with $\sigma^2=8$ (violet curve), to the spectrum reconstructed from Eq.~\eqref{xasreconstruction}, where the total spectrum (black line) has been rescaled with an overall factor. The reconstruction of the trXAS signal from the nonequilibrium density matrix works relatively well, which indicates that XAS can be used to determine the dynamics of the local occupations. 

Of course there are deviations from the atomic picture. On the one hand, the bare exciton lines experience small shifts with respect to the full DMFT result, which can be understood as hybridization shifts, just like in the equilibrium spectra. More importantly, for the interpretation one should take into account that in the nonequilibrium situation there are additional effects on the lineshape and position of the exciton peaks:

(i) Initial state effects: The initial states $\alpha=1_d$ and $\alpha=3_d$ are delocalized singlons and triplons, and their energy should therefore be taken from a broader distribution which is determined by the electron addition and removal spectrum, respectively, i.e., the unoccupied density of states within the lower Hubbard band (singlons) and the occupied density of states in the upper Hubbard band (triplons). Previous nonequilibrium DMFT simulations have shown that the initial energy distribution of excited singlons and triplons after an impulsive excitation covers almost the entire upper and lower Hubbard band,\cite{Eckstein2011} but as these excitations lose kinetic energy by creating additional Hund's excitations $2_s$ and $2_l$, the singlon and triplon distribution relaxes towards the upper edge of the lower Hubbard band and lower edge of the upper Hubbard band, respectively.\cite{Strand2017}  Hence one might expect a corresponding time-dependent shift and narrowing of the signals associated with the transitions $1_d\to2_x$ and $3_d\to4_d$. However, in the present case this initial relaxation dynamics happens within a few hopping times (see the initial increase of $2_s$ and $2_l$ states in Fig.~\ref{fig_xas_half_states}), and 
therefore cannot be resolved 
given the restrictions imposed by the energy-time uncertainty of the probe pulse. 
The remaining effect of the initial state delocalization is therefore an additional broadening, which is consistent with the data.

(ii) Final state effects and additional hybridization with the continuum: In the nonequilibrium state, there are additional   continuum excitations compared to the initial equilibrium state. More specifically, for each exciton $\underline \alpha$, we can look for states $\underline \alpha' - \beta_i + \beta$, where the local state is transformed into $\underline \alpha'$ with a different $d$ occupation, while a state in the environment is switched from $\beta_i$ to $\beta$. In equilibrium, only $\beta_i=2_h$ needs to be considered, but in the excited state there are many more possibilities. For example, the $\underline 3_d$ exciton can hybridize with a doublon  exciton $\underline 2_x$ ($x=h,l,s$) by a transition which involves the decay of a photo-excited singlon, $\beta_i=1_d\to\beta=2_x$. Whether the initial exciton $\underline \alpha$ is well-defined will depend on whether or not its excitation energy overlaps with the continuum. Table~\ref{tabstates} gives an overview over the possible processes for the present case. The first column is the initial state configuration,  the second columns show accessible exciton excitations $\underline{\alpha}$, with the given excitation energy $E_\text{ex}=E(\underline{\alpha})-E(\alpha_i)$. The subsequent columns  show continuum excitations which are accessible from the exciton $\underline{\alpha}$ by a single particle exchange with the continuum; only those continuum states are shown which are in close resonance with the exciton, i.e., with energy $E_\text{cont}$ such that  $|\Delta E| = |E_\text{cont}-E_\text{ex}| < 3$, sufficiently smaller than the width of the Hubbard bands. One can see that the $\underline 1_d$ and $\underline 4_d$ excitons are already close to a continuum state if the environment is in equilibrium (their decay involves a $\beta_i=2_h$ state). 
For the other excitons, decay channels are activated by nonequilibrium $1_d$ and $3_d$ populations, which is in agreement with the observation that all exciton lines are broadened with respect to the atomic line-width.

\begin{figure}[t]
\begin{center}
\includegraphics[angle=-90, width=0.45\textwidth]{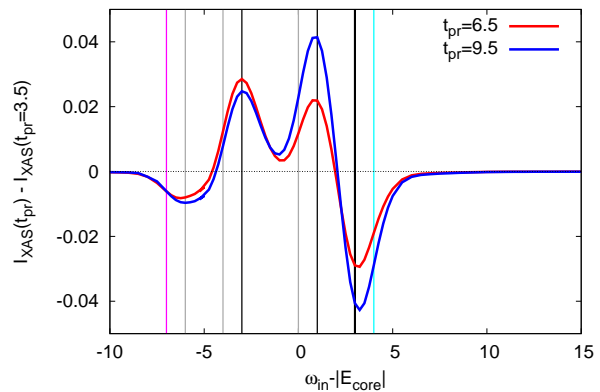}
\caption{
Time evolution of the XAS spectrum for $U=10$, $J=2$, $U_{cd}=3$, and $v_{\rm pu}=1$. The red curve plots the difference 
$I_{\rm XAS}(t_{\rm pr}=6.5)-I_{\rm XAS}(t_{\rm pr}=3.5)$ and the blue curve the difference $I_{\rm XAS}(t_{\rm pr}=9.5)-I_{\rm XAS}(t_{\rm pr}=3.5)$.  
Here, a short probe pulse with $\sigma^2=2$ is used. The vertical lines indicate the energies associated with the following atomic transitions: $2_h\rightarrow \underline{3}_d$ (thick black), $2_{l,s}\rightarrow \underline{3}_d$ (thin black), $1_d\rightarrow \underline{2}_x$ (gray), $3_d\rightarrow \underline{4}_d$ (light blue), $0_d\rightarrow \underline{1}_d$ (pink).
}
\label{fig_xas_half_time}
\end{center}
\end{figure}   

It follows from the previous discussion that the time evolution of the XAS signal allows to track the change in the population of the different local states. Because the initial dynamics during the pump is so fast that the individual multiplet peaks are difficult to resolve given the energy-time uncertainly of a sufficiently short pulse,\cite{Strand2017,Gillmeister2020} we focus on the dynamics after the pump. Figure~\ref{fig_xas_half_time} plots the difference in the signal at $t_{\rm pr}=6.5$ and $t_{\rm pr}=3.5$ (red curve) as well as $t_{\rm pr}=9.5$ and $t_{\rm pr}=3.5$ (blue curve), measured with the short probe pulse ($\sigma^2=2$).  These curves in particular show the increasing population of the $2_l\to\underline 3_d$ exciton (black vertical line at $\omega_\text{in}-|E_\text{core}|=1$), which is consistent with the increase of the $2_l$ occupation  for times $3.5<t<9.5$ (green line in Fig.~\ref{fig_xas_half_states}). They also reveal an initially strong 
increase followed by a slow decrease
of the $2_s\to\underline3_d$ exciton (black vertical line at $\omega_\text{in}-|E_\text{core}|=-3$), which tracks the behavior of the $2_s$ occupation (dark blue line in Fig.~\ref{fig_xas_half_states}). Similarly, one can see clear signatures of the decrease in the high-spin doublon $2_h$ and triplon states $3_d$ states (red and pink lines in Fig.~\ref{fig_xas_half_states}; thick black and light blue vertical lines in Fig.~\ref{fig_xas_half_time}) and
in the singlon states $1_d$ (same as pink line in Fig.~\ref{fig_xas_half_states}; gray vertical lines, especially at $\omega_\text{in}-|E_\text{core}|=-6$ in Fig.~\ref{fig_xas_half_time}).

\subsection{Quarter-filled system} 
\label{sec_quarter}

\subsubsection{Spectral functions}

In this section, we consider the quarter-filled two-band model with $U=10$, $J=2$, 
whose equilibrium DOS is very different from the half-filled case, as one can see by comparing Fig.~\ref{fig_half}(a) and Fig.~\ref{fig_quarter}(a). Here, we place the core level at $E_\text{core}=-20$, and the highest energy subband of the upper Hubbard band (associated with the creation of $2_s$) at energy  
$E_{2_s}=8$. 
This gives two conditions for the electron addition energy, $E(2_s)-E(1_d)=E_\text{int}(2_s)-\mu+2U_{cd}\equiv E_{2_s}$, and for the core electron removal energy, $-[E(\underline{1}_d)-E(1_d)]=\epsilon_c-\mu+U_{cd}\equiv E_\text{core}$, which determine the values of $\mu$ and $\epsilon_c$: $\mu=-E_{2_s}+E_\text{int}(2_s)+2U_{cd}$, $\epsilon_c=E_\text{core}-E_{2_s}+E_\text{int}(2_s)+U_{cd}$. 
With these parameters, the lower Hubbard band, which represents singly occupied states, is centered at $\omega \approx -2$, and the subbands of the upper Hubbard band are found near $\omega \approx 8$ ($\equiv E_{2_s}$), $4$ $(\equiv E_{2_l})$, and $2$ $(\equiv E_{2_h})$, as shown in Fig.~\ref{fig_quarter}(a). From high to low energies, these subbands are associated with the creation of the states $2_s$, $2_l$ and $2_h$.
As in the half-filled case, the spectral function of the core level is broadened by the coupling to a bath with a box-shaped density of states and bandwidth $W=20$ (dashed black line in Fig.~\ref{fig_quarter}(a)). The coupling strength to this bath is $v_\text{bath}=1$.

\subsubsection{XAS in equilibrium}

\begin{figure}[t]
\begin{center}
\includegraphics[angle=-90, width=0.45\textwidth]{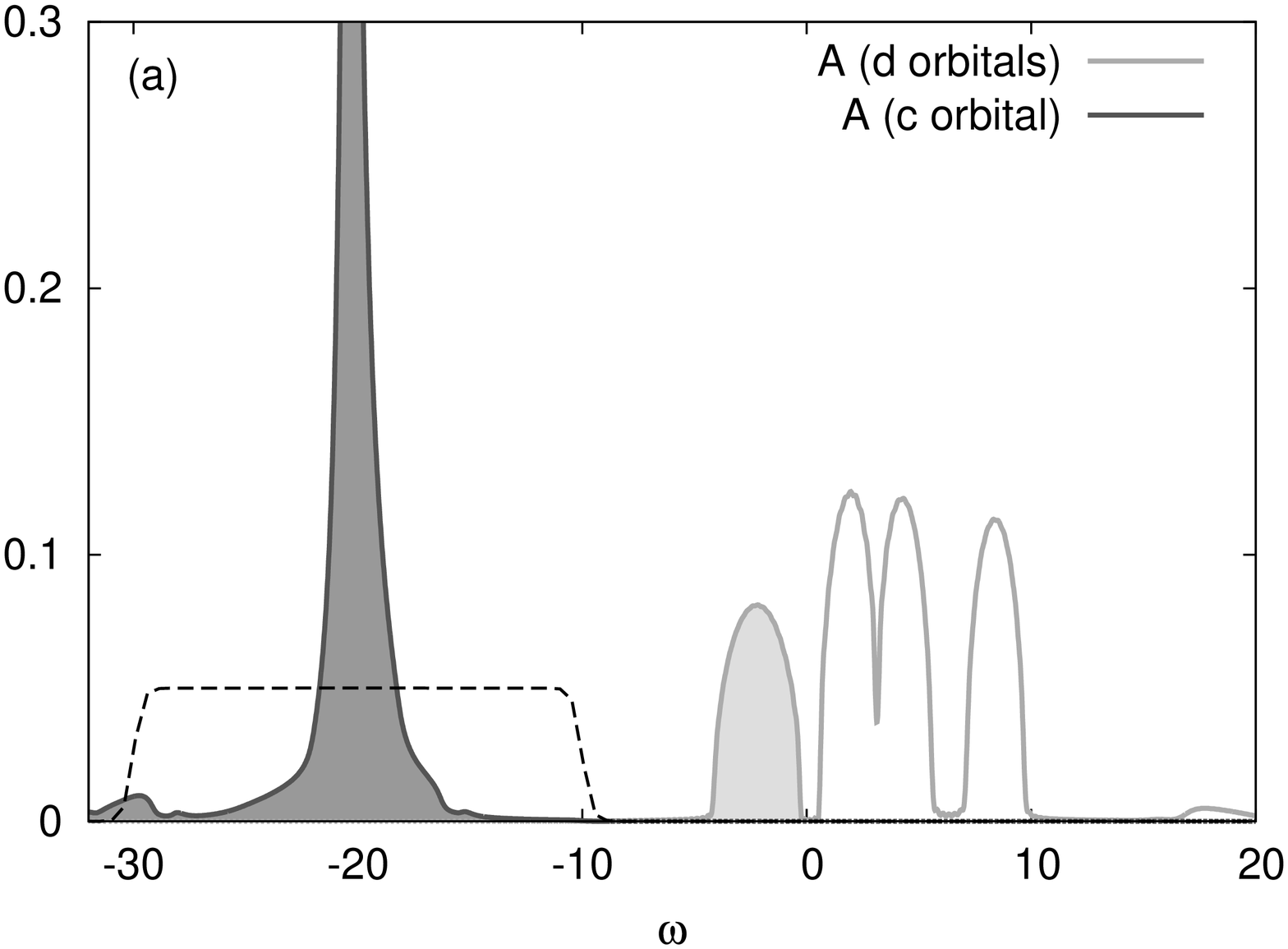}\\
\hspace{0mm}
\includegraphics[angle=-90, width=0.44\textwidth]{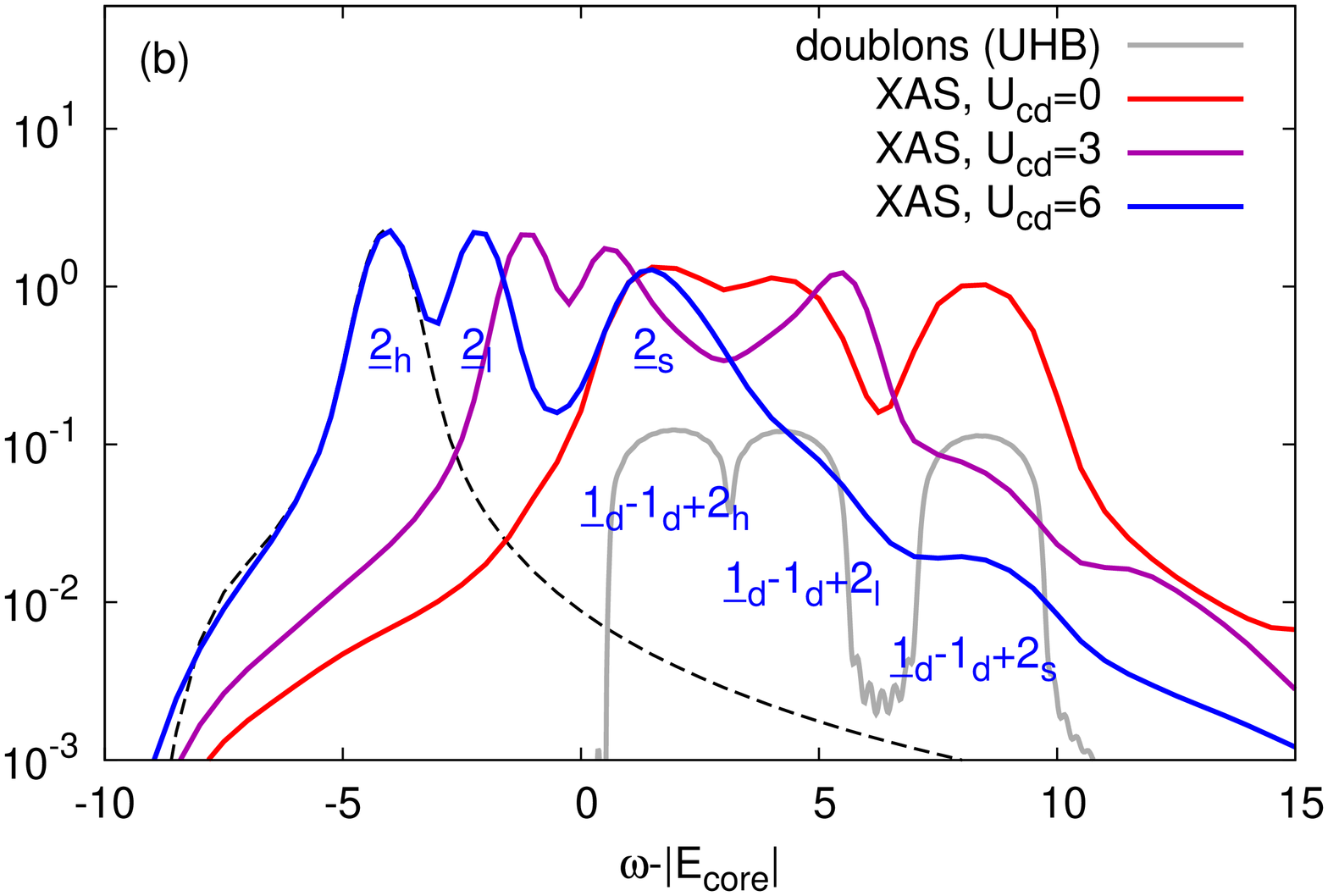}
\caption{
(a) Equilibrium spectral functions of the quarter-filled model with $U=10$, $J=2$, 
and
$E_\text{core}=-20$. The coupling to a Fermion bath with box-shaped DOS with bandwidth $W=20$ (dashed black line) and coupling strength $v_\text{bath}=1$  
leads to a broadening of the $c$-DOS. The lower Hubbard band (singly occupied sites) in the $d$-DOS is centered near $\omega=-2$, while the upper Hubbard band (doubly occupied sites) features three subbands centered near $\omega=2$, $4$ and $8$.  
(b) Equilibrium XAS spectrum for the indicated values of $U_{cd}$ and a probe pulse with $\sigma^2=8$. For comparison, we also show in gray the shifted upper Hubbard band (states $\underline{1}_d-1_d+2_x$). 
}
\label{fig_quarter}
\end{center}
\end{figure}

The equilibrium XAS signals for different values of $U_{cd}$ are plotted in Fig.~\ref{fig_quarter}(b). In the atomic limit, the spectrum features three peaks associated with the transitions $1_d\rightarrow \underline 2_x$ ($x=h,l,s$). The corresponding energies are $E(\underline 2_x)-E(1_d)-|E_\text{core}|=E_\text{int}({2_x})+E_{2_s}-E_\text{int}(2_s)-U_{cd}=E_{2_x}-U_{cd}$. With increasing core hole interaction, the XAS peaks are shifted to lower energy by $E_\text{bind}=-U_{cd}$.  
In the lattice system, for large enough $U_{cd}$, we expect exciton features near these energies, with a lineshape similar to that of the atomic system (see dashed black line for the $\underline 2_h$ exciton peak for $U_{cd}=6$). In addition, there can be continuum features associated with the hopping of the doublon from the site with core hole into the rest of the lattice, $\underline 2_x\rightarrow \underline 1_d-1_d+2_y$. Since the energy difference between the initial and final state 
in this decay 
is $U_{cd}+E_\text{int}(2_y)-E_\text{int}(2_x)$, it is clear that for $U_{cd}=0$, the continuum features which do not involve any change in the doublon configuration ($x=y$) appear at the same energies as the exciton. We thus obtain three strongly broadened peaks which overlap with the appropriately shifted upper Hubbard band of the spectral function (gray line in Fig.~\ref{fig_quarter}(b)). 

For $U_{cd}=3$, the lowest-energy exciton peak exhibits little overlap with the continuum and becomes sharp, while the middle and high-energy peaks are broadened because they overlap with continuum features associated with $x\ne y$. For $U_{cd}=6$, only the highest-energy peak is affected by such a broadening. The other continuum features are evident as humps near the energies of the shifted upper Hubbard band, as indicated by the blue labels. 

\begin{figure}[t]
\begin{center}
\includegraphics[angle=-90, width=0.45\textwidth]{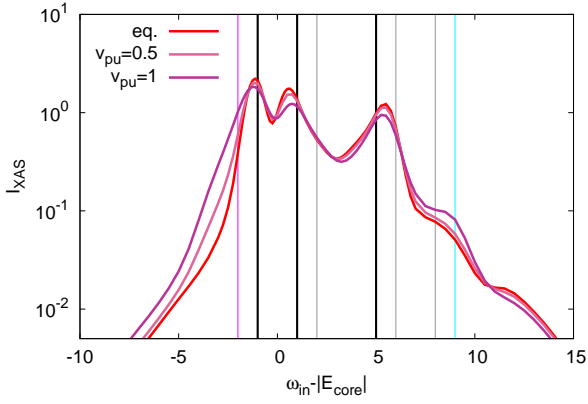} 
\caption{
XAS spectra of the equilibrium and photo-doped quarter-filled model with $U=10$, $J=2$, and $U_{cd}=3$. We use pump pulses with $v_{\rm pu}=0.5$ and $1$, pump pulse frequency $\omega_{\rm pu}=U$ (resonant excitation to the highest-energy subband $E_{2_s}$) and long probe pulses with $\sigma^2=8$ centered at $t_\text{pr}=6.5$. The vertical lines indicate the energies associated with the following atomic transitions:  $1_d\rightarrow \underline{2}_x$ (thick black), $0_{d}\rightarrow \underline{1}_d$ (pink), $2_x\rightarrow \underline{3}_d$ (gray), $3_d\rightarrow \underline{4}_d$ (light blue). 
}
\label{fig_xas_quarter}
\end{center}
\end{figure}   

\subsubsection{XAS after photo-doping}

The XAS signals after photo-excitation with $\omega_{\rm pu}=U=10$ are shown in Fig.~\ref{fig_xas_quarter} 
for $U_{cd}=3$ and compared to the equilibrium result (red line). Here, the pump pulse envelope is the same as in Fig.~\ref{fig_xas_half_states}(a) and the probe pulse with $\sigma^2=8$ is centered at $t_\text{pr}=6.5$. 
The black vertical lines at $\omega_\text{in}-|E_\text{core}|=-1$, $1$, $5$, which correspond to the final state excitons $\underline 2_h$, $\underline 2_l$, $\underline 2_s$, respectively, show that excitations associated with the creation of doublon states ($1_d\rightarrow \underline 2_x$) indeed explain the main equilibrium XAS features (red line). 
As in the half filled case, additional local states should contribute to the XAS signal after the photo-excitation. 
Transitions associated with empty states ($0_d \rightarrow \underline{1}_d$) should produce a peak near
$\omega_\text{in}-|E_\text{core}|=-2$ (pink line). Transitions associated with the three different doublon states ($2_x\rightarrow \underline{3}_d$) should result in three additional features at higher energies: for the $2_s$, $2_l$, $2_h$ doublons, we expect them to appear near $\omega_\text{in}-|E_\text{core}|=2$, $6$, $8$, respectively (gray lines). Core excitations to triplons ($3_d\rightarrow \underline{4}_d$) should produce a single peak near  $\omega_\text{in}-|E_\text{core}|=9$ (light blue line). 

\begin{figure}[t]
\begin{center} 
\includegraphics[angle=-90, width=0.45\textwidth]{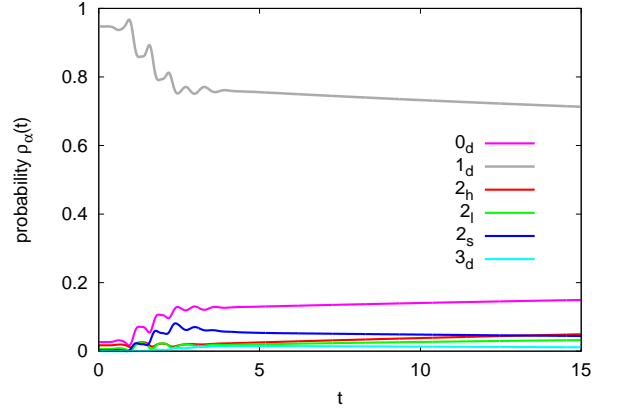}
\caption{
Probability of different local states in the quarter-filled model with $U=10$, $J=2$, $U_{cd}=3$, 
which is excited by a pump pulse with $v_{\rm pu}=1$ and $\omega_{\rm pu}=10$. The pulse envelope is the same as shown in Fig.~\ref{fig_xas_half_states}(a). 
}
\label{fig_xas_quarter_states}
\end{center}
\end{figure}   

The XAS signals after photo-excitation with $\omega_{\rm pu}=U$ and $v_\text{pu}=0.5$, $1$ are shown in Fig.~\ref{fig_xas_quarter} by the pink-violet lines. 
We observe some broadenings and shifts of the main XAS peaks relative to the equilibrium spectrum, as well as the enhancement of some of the satellite features. In the energy interval $-2\lesssim \omega_\text{in}-|E_\text{core}|\lesssim 6$, the enhancement of the signal near 
the exciton transitions starting from the exited states (empty initial states $0_d$ shown by the pink vertical line and doubly occupied states $2_x$  shown by grey vertical lines)  is partly compensated or even overcompensated by the reduction of the peak amplitudes near the exciton transitions starting from the dominant equilibrium state $1_d$ (black lines). This results in 
a reduced signal and small energy shifts of the peaks. The enhancement of the signal associated with $2_h$ doublons and triplons is however clearly seen at higher energies. 

Overall, the relative changes in the XAS signal are smaller than in the half-filled case (Fig.~\ref{fig_xas_half}), which can be explained by the weaker absorption of the quarter-filled system. 
The evolution of the local states after the pulse with $v_{\rm pu}=1$ is plotted in Fig.~\ref{fig_xas_quarter_states} and indeed shows smaller relative changes of the different local state populations than observed in the half-filled system (Fig.~\ref{fig_xas_half_states}(b)). Apart from the expected decrease in the initially dominating $1_d$ configurations, this figure demonstrates a strong increase in the $0_d$ population, which roughly matches the total increase in the $2_x$ populations. This is expected, since the pump pulse primarily converts two singlons on neighboring sites into a doublon and an empty site. Among the $2_x$ populations, we find the strongest increase for $2_s$ during the pump pulse, because $\omega_\text{pu}=U$ mainly excites electrons into the highest-energy subband of the upper Hubbard band. In the dynamics after the pump, we observe the conversion of high-energy doublons $2_s$ into lower-energy doublons $2_l$ and $2_h$. We also  note that the conversion of a $0_d$ state and a neighboring $2_s$ state into an empty site and a triplon costs an energy $U$, which matches the energy released by the decay of $2_s$ into two singlons. Hence, if the pulse creates a substantial density of $2_s$ and $0_d$ states, these can trigger the production of triplons. After the current pump pulse, these densities are however so low that the triplon population increases only slowly.  

\begin{figure}[t]
\begin{center}
\includegraphics[angle=0, width=0.5\textwidth]{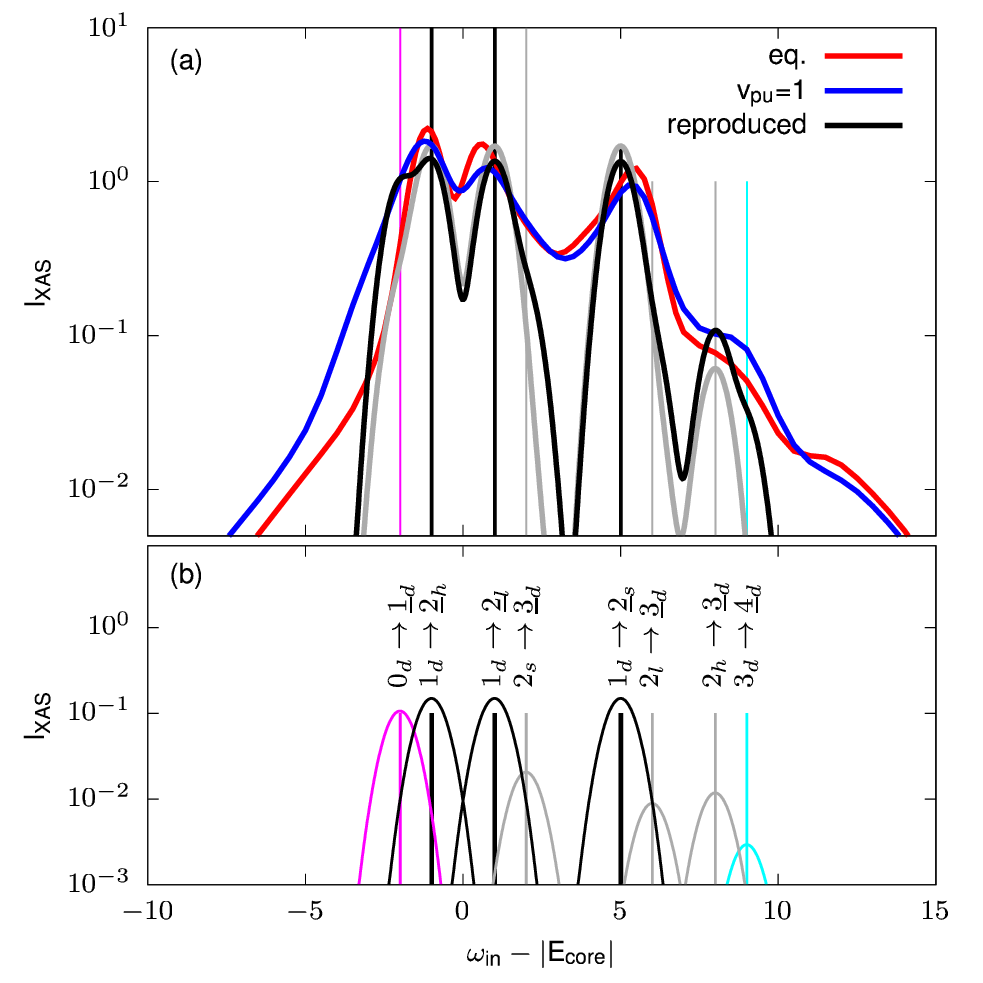} 
\caption{
(a) XAS spectra of the photo-doped quarter-filled model with $U=10$, $J=2$, $U_{cd}=3$, $v_{\rm pu}=1$, $\omega_{\rm pu}=10$. The results are obtained at probe time $t_{\rm pr}=6.5$ with the long probe pulse ($\sigma^2=8$). 
The full result (blue line) is compared to the XAS spectrum reconstructed from the atomic system with the nonequilibrium occupations measured at $t_\text{pr}=6.5$ (black line). The gray line is the reproduced spectrum of the initial state. The vertical lines indicate the energies associated with the following atomic transitions: $1_d\rightarrow \underline{2}_x$ (thick black), $0_{d}\rightarrow \underline{1}_d$ (pink), $2_x\rightarrow \underline{3}_d$ (gray), $3_d\rightarrow \underline{4}_d$ (light blue), as indicated in panel (b), which also shows the contributions of the different excitons to the spectrum. For the reconstruction, we use the lineshape in Eq.~\eqref{atomicline} with $\sigma_\text{eff}^2=5.5$, as measured in the atomic limit.
}
\label{fig_xas_quarter_reproduced}
\end{center}
\end{figure}   
 
 As in the half-filled case,  it is an interesting question how well the trXAS spectrum can be reproduced with broadened atomic peaks, whose weights are chosen according to the nonequilibrium distribution $\rho_\alpha(t)$ plotted in Fig.~\ref{fig_xas_quarter_states}. The result for $v_\text{pu}=1$ and $t_{\rm pr}=6.5$ is shown in Fig.~\ref{fig_xas_quarter_reproduced}, which compares the nonequilibrium DMFT result (blue line) to the reconstructed spectrum (black line). In addition, we  also show here the initial equilibrium spectrum (red line) and the reconstructed spectrum of the initial equilibrium state (gray line). The matrix elements listed in Table~\ref{tabstates} have been used to convert the valence occupations into the weights of the different atomic peaks, and we again employ Gaussians with $\sigma^2_\text{eff}=5.5$ (Appendix~\ref{app:tests}) to reconstruct the nonequilibrium XAS signal. 
The comparison with the full DMFT result shows that the actual broadening is significantly larger 
than in the half-filled case analyzed in Fig.~\ref{fig_xas_reproduced}. This is because 
the high-energy subband overlaps with the continuum (Fig.~\ref{fig_quarter}(b)), and also several of the photo-induced features are energetically close to continuum states (Table~\ref{tabstates}). 
(Note that while for the quarter filled system the positions of the excitonic transitions are different from the half-filled system, because the values for $\mu$ and $\epsilon_c$ differ, the energy {\em difference} between the continuum and the excitons is independent of $\mu$ and $\epsilon_c$ and therefore  the same as listed in Table~\ref{tabstates}.)
Many of the near-resonant exciton decay paths for $U_{cd}=3$ involve singly occupied initial states on the lattice (continuum states with ``$-1_d$'' in the table) and are therefore active already after weak excitations, while for the half-filled system the $1_d$ singlons first need to be populated by the pulse. 
Because the resulting broadening effect is missing in the reconstructed spectrum,  
 a better agreement with the nonequilibrium DMFT result would be obtained for $\sigma^2_\text{eff}\approx 3$. 
Nevertheless, at the qualitative level, we can see that the change in the reconstructed spectra (gray and black lines) correctly reproduces the photo-induced change in the DMFT results. In particular, it captures the significant broadening and shift of the lowest-energy peak, which comes from the enhanced population of empty states, and the significant relative increase of the shoulder structure around $\omega_\text{in}-|E_\text{core}|\approx 8$-$9$. It also correctly describes the reduction of the main peaks near the black vertical lines, and the slight shift of the dominant peak near $\omega_\text{in}-|E_\text{core}|\approx 5$-$6$ to higher energy. 

\begin{figure}[t]
\begin{center}
\includegraphics[angle=-90, width=0.45\textwidth]{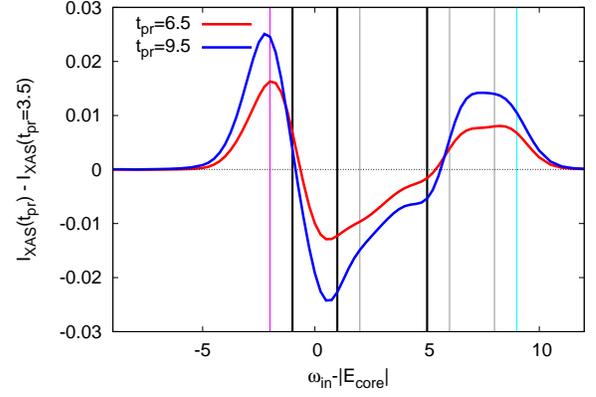} 
\caption{
Time evolution of the XAS spectrum in the quarter-filled model with $U=10$, $J=2$, $U_{cd}=3$ and $v_{\rm pu}=1$. The red curve shows the difference 
$I_{\rm XAS}(t_{\rm pr}=6.5)-I_{\rm XAS}(t_{\rm pr}=3.5)$ and the blue curve the difference $I_{\rm XAS}(t_{\rm pr}=9.5)-I_{\rm XAS}(t_{\rm pr}=3.5)$.  
The vertical lines indicate the energies associated with the following atomic transitions: $1_d\rightarrow \underline{2}_x$ (thick black), $0_{d}\rightarrow \underline{1}_d$ (pink), $2_x\rightarrow \underline{3}_d$ (gray), $3_d\rightarrow \underline{4}_d$ (light blue). Here, we use the short probe pulse with $\sigma^2=2$. 
}
\label{fig_xas_quarter_time}
\end{center}
\end{figure}   

As in the case of the half-filled system, the evolution in the population of the local states is reflected in the trXAS signal. To illustrate this, we plot in Fig.~\ref{fig_xas_quarter_time} the differences $I_
\text{XAS}(t_{\rm pr}=9.5)-I_\text{XAS}(t_{\rm pr}=3.5)$ (red) and $I_\text{XAS}(t_{\rm pr}=6.5)-I_\text{XAS}(t_{\rm pr}=3.5)$ (blue). For these calculations, we use the shorter probe pulse with $\sigma^2=2$.  
One notices the increase in the signal associated with the triplons (pink line) and $2_h$ doublons (gray line at $\omega_\text{in}-|E_\text{core}|=8$), which matches the evolution of the corresponding probabilities $\rho_\alpha(t)$ after the pump (Fig.~\ref{fig_xas_quarter_states}). We further see a decrease in the signals near the black vertical lines (except near $\omega_\text{in}-|E_\text{core}|=-1$, where this decrease is overcompensated by the increase in the triplon signal) and the lowest-energy gray line, which is consistent with the decrease of the $1_d$ and $2_s$ populations after the pump. The kink-like structure near $\omega_\text{in}-|E_\text{core}|=5$ reflects the decrease in the $1_d$ population and simultaneous increase in the $2_l$ population.

\subsubsection{Effect of the pump frequency}

\begin{figure}[t]
\begin{center}
\includegraphics[angle=-90, width=0.45\textwidth]{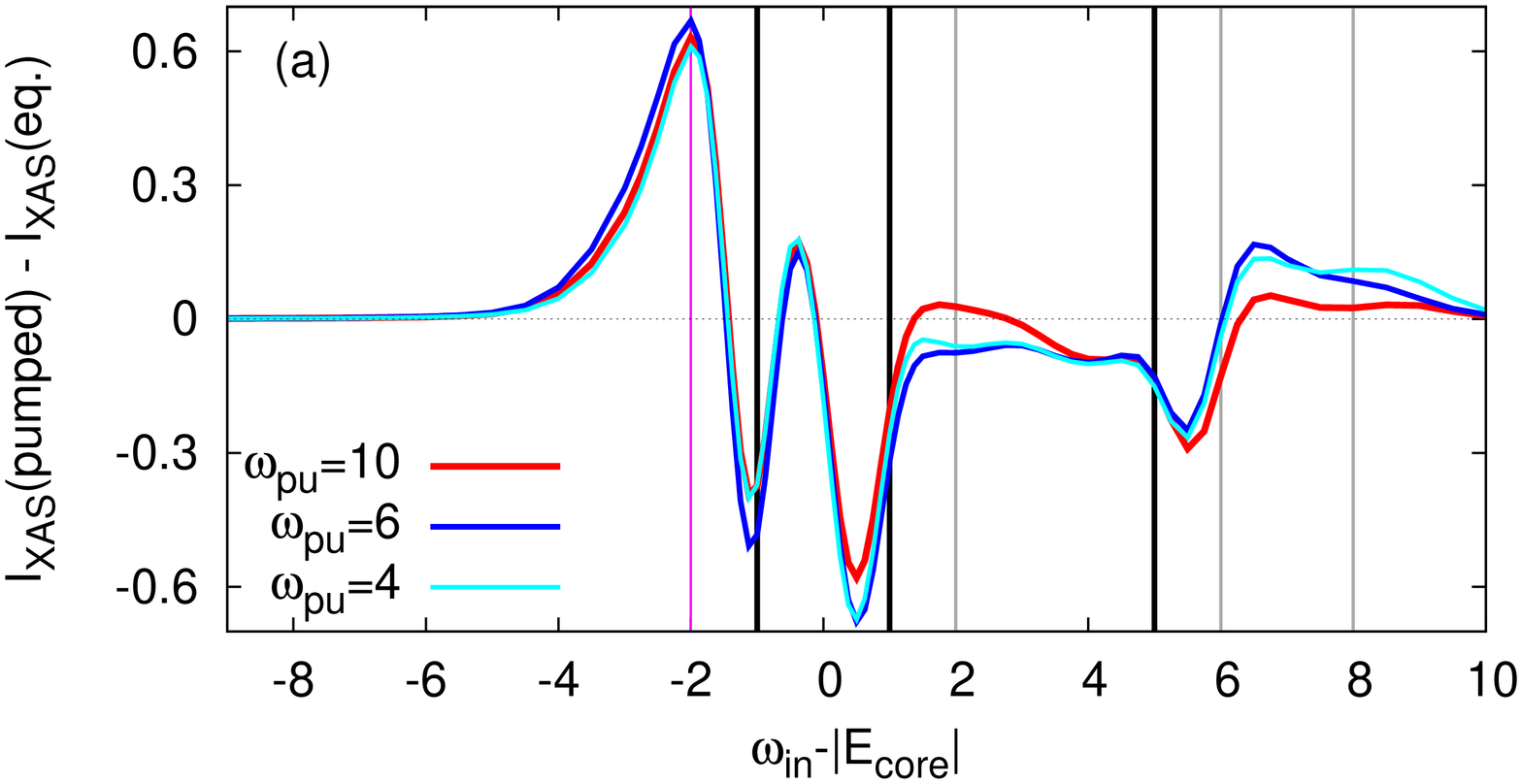} 
\includegraphics[angle=-90, width=0.45\textwidth]{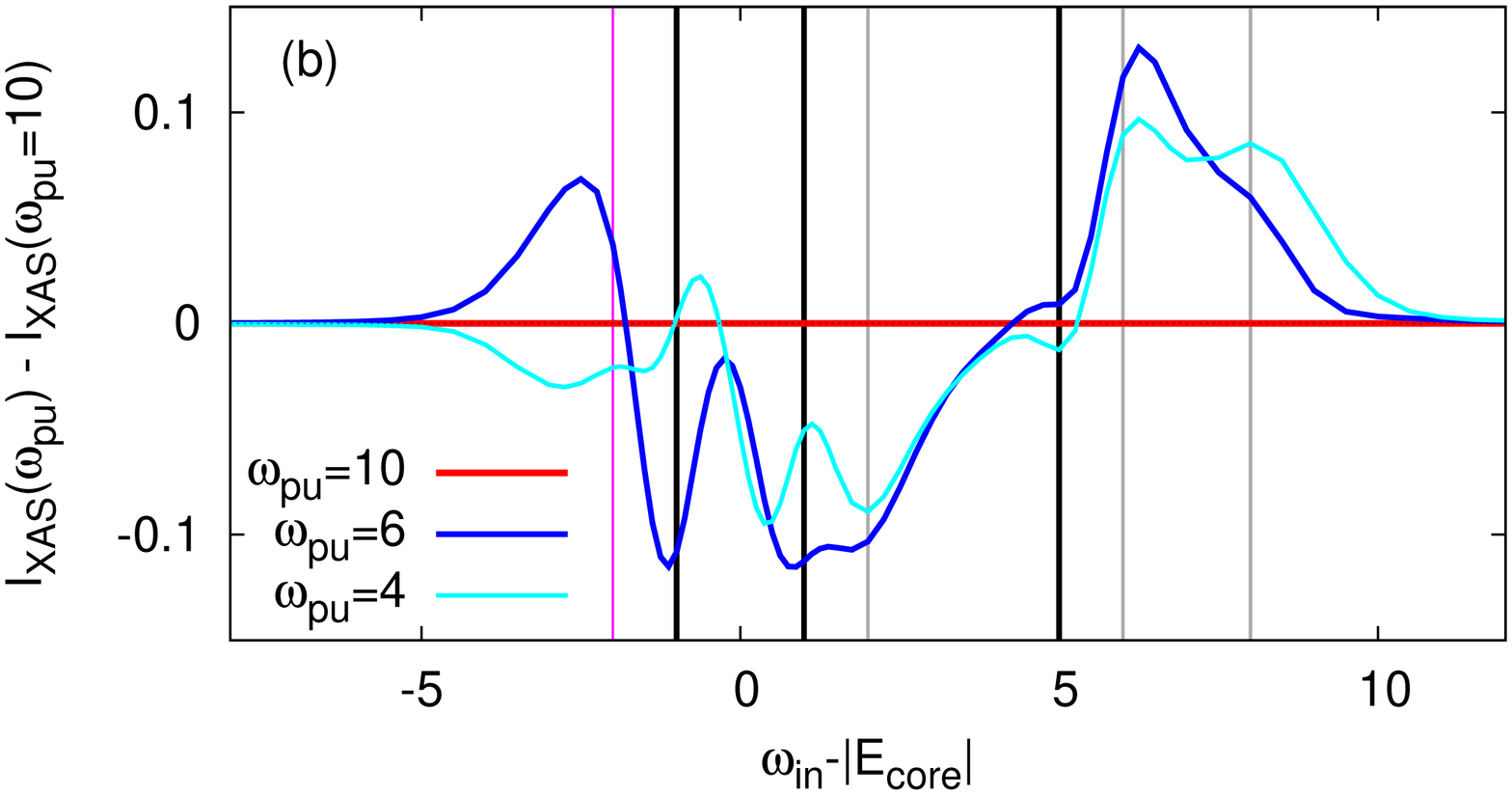} 
\caption{
Pulse frequency dependence of the nonequilibrium XAS spectrum in the quarter-filled model with $U=10$, $J=2$, and $U_{cd}=3$. The pulse has amplitude $v_{\rm pu}=1$ and we consider the three different resonant excitations: $\omega_{\rm pu}=10$, $6$, $4$. In panel (a), the curves show the difference $I_\text{XAS}$(pumped)$-I_\text{XAS}$(eq.) and in panel (b) $I_\text{XAS}$($\omega_\text{pu}$)$-I_\text{XAS}$($\omega_\text{pu}=10$). 
The vertical lines indicate the energies associated with the following atomic transitions: $1_d\rightarrow \underline{2}_x$ (thick black), $0_{d}\rightarrow \underline{1}_d$ (pink), $2_x\rightarrow \underline{3}_d$ (gray), $3_d\rightarrow \underline{4}_d$ (light blue). Here, we use the long probe pulse with $\sigma^2=8$ centered at $t_\text{pr}=6.5$. 
}
\label{fig_xas_quarter_w_diff}
\end{center}
\end{figure}

Since the upper Hubbard band of the quarter-filled system splits into three subbands, one might expect that the trXAS signal exhibits a characteristic dependence on the frequency of the pump pulse, which could be exploited to identify the origin of the different spectral features. 
In Fig.~\ref{fig_xas_quarter_w_diff}, we plot measurements for the three different resonant excitations to these subbands at $E_{2_s}$, $E_{2_l}$, $E_{2_h}$:  $\omega_{\rm pu}=10$, $6$, $4$. 
Panel (a) shows the difference between the photo-doped XAS signal (measured at $t_\text{pr}=6.5$ with the long probe pulse with $\sigma^2=8$) and the equilibrium spectrum. 
The expected enhancement of the $2_s$, $2_l$ and $2_h$ populations by the three different pulses 
is clearly reflected in the data, which show an increase near the corresponding gray lines with $\omega_\text{in}-|E_\text{core}|=2$, $6$, $8$, respectively. 
(For the gray lines next to black ones, the increase is partially masked by the decrease in the $1_d$ population, which results in a kink-like structure.)
The evolution of the XAS signal also demonstrates an increase in the empty site population after all these pulses (peaks near the vertical pink line), which is expected since all three pulses convert singlons into doublons and empty sites. 

In panel (b), we plot the difference between the nonequilibrium spectra for $\omega_\text{pu}=4$, $6$ and the one for $\omega_\text{pu}=10$. This figure shows more clearly that the $\omega_\text{pu}=4$ pulse creates more high-spin doublons $2_h$ relative to the $\omega_\text{pu}=10$ pulse (maximum in the light-blue line near $\omega_\text{in}-|E_\text{core}|=8$), and less $2_s$ doublons (minimum near $\omega_\text{in}-|E_\text{core}|=2$). On the other hand, the $\omega_\text{pu}=6$ pulse creates more $2_l$ doublons (maximum in the dark-blue line near $\omega_\text{in}-|E_\text{core}|=6$) and also less $2_s$ doublons (minimum near $\omega_\text{in}-|E_\text{core}|=2$). 
The data around the black line with $\omega_\text{in}-|E_\text{core}|=-1$ furthermore show that the $\omega_\text{pu}=6$ pulse removes $1_d$ states more efficiently than the other two pulses. This is because of the slightly larger total absorption for $\omega_\text{pu}=6$, which excites core electrons to the center of the absorption band.

 \section{Conclusions}
 \label{sec:conclusions}
We proposed a formalism to compute the dynamics of time-resolved XAS based on nonequilibrium DMFT and applied the procedure to photo-doped Mott insulators. The calculation is performed in two stages. First, we simulate the dynamics in the valence bands after a photo-excitation using DMFT. Second, the XAS signal is extracted by the explicit treatment of core orbitals in an extended impurity problem and time-dependent simulations of core-valence excitations. The equilibrium XAS signal is dominated by excitonic resonances between a core hole and valence electrons in the upper Hubbard band due to their strong interaction, while the weaker continuum features are associated with delocalized states. After a photo-excitation, the non-thermal population of local many-body states activates new excitonic features in XAS and modifies the lifetime of resonances due to additional decay channels. We calculated these effects by considering the two-band Hubbard model at half- and quarter-filling and connected the photo-induced XAS features to the activated transitions to different multiplets. The analysis exemplifies how a simple pump-probe protocol combined with XAS allows us to determine 
the time-dependent population of different local states
in correlated solids and complements previous analyses based on time-resolved photo-emission~\cite{Strand2017,Gillmeister2020,ligges2018} or strong-field effects.\cite{Dasari2020}

In previous studies, the equilibrium XAS signal has been successfully analyzed with small cluster calculations.\cite{derLann1986,Haverkort2012} An important outcome of our time-dependent calculation is the demonstration that such cluster calculations can be generalized to time-resolved XAS by properly adjusting the nonthermal multiplet population. Such an analysis will be particularly useful for describing realistic correlated solids, because the high computational complexity of explicit time-dependent calculations imposes severe constraints.\cite{Wang2020} 

Despite the similarities between the two approaches, the time-dependent DMFT formalism offers some significant benefits. In particular, small cluster calculations cannot capture the broadening associated with the decay into continuum excitations, and one may expect severe finite-size effects, because photo-induced charge carriers are mobile. Since the DMFT bath represents the infinite environment of the probed site, our nonequilibrium DMFT method does not suffer from these limitations. Moreover, by following the dynamics of the photo-induced state population we can learn how highly excited charge carriers are coupled to bosonic degrees of freedom, including orbital, antiferromagnetic, or lattice excitations. A significant technical challenge is extending the analysis to longer probe times (higher energy resolution), which will shed light on these couplings via photo-induced peak shifts and lifetime changes in trXAS.

Extending the modeling beyond on-site Hubbard interactions will allow us to follow the dynamics of crystal-field excitations in charge-transfer insulators or explore the vibrational control of $d$-$d$ excitations by infra-red pump pulses.\cite{Marciniak2021} Finally, the presence of excitons in XAS leads to a high sensitivity to photo-induced changes in the screening, whose dynamics is essential to understand the photo-manipulation of bandgaps in correlated solids.\cite{Golez2019,Baykusheva2022}

\acknowledgments
P.W. acknowledges support from ERC Consolidator Grant 724103 and SNSF Grant No. 200021-196966, and  M.E. from ERC Starting Grant No.~716648. DG is supported by Slovenian Research Agency
(ARRS) under Program J1-2455 and P1-0044. The calculations were run on the Beo05 cluster at the University of Fribourg.

\appendix

\section{Rotating wave approximation}
\label{app:rotwave}

Usually, there is a large energy  scale separation between the bandwidth of the valence band, which is at most of the order of a few eV, and the energies $\omega_{\text{\text{in}}}$, and $|\epsilon_{c}|$, which can be of the order of  keV.  In this case, the numerical evaluation can be simplified using a rotating wave approximation. For completeness, we summarize the corresponding equations in this appendix. 

We assume that $\omega_{\text{in}}=\tilde \omega_{\text{in}}+E$,  $\epsilon_c=-E+\tilde \epsilon_c$, with $\tilde \omega_{\text{in}},\tilde\epsilon_c\ll E$. To make use of a rotating wave approximation, the Hamiltonian $H_\text{loc}$  [Eq.~\eqref{hloc}]  is then rewritten using a canonical transformation which shifts the energy of the core band. In general, after a time-dependent basis transformation  $|\tilde \psi\rangle = \mathcal{W}(t) |\psi\rangle$ the new wavefunction satisfies the Schr\"odinger equation $i\partial_t |\tilde \psi\rangle = \tilde H| \tilde \psi\rangle$ with $\tilde H=\mathcal{W}H\mathcal{W}^\dagger +i(\partial_t \mathcal{W})\mathcal{W}^\dagger$. The choice $\mathcal{W}=\exp\big[-itE\sum_{\sigma}c_{\sigma}^\dagger c_{\sigma} \big]$ 
leads to 
\begin{align}
\label{hloc_RWA}
\tilde H_\text{loc} 
&=
\tilde H_d 
+
\tilde H_c
+
\tilde H_{cd}
+
\tilde H_{\text{\text{dip}}},
\end{align}
where the valence Hamiltonian $H_{d}$ and  the core-valence interaction $H_{cd}$ remain unchanged,
but the core energy in $\tilde H_{c}$ is shifted to $\tilde \epsilon_{c}$. The dipolar Hamiltonian becomes
\begin{align}
\label{ingoung}
&\tilde H_{\text{dip}}
=
\big[s(t) e^{-i\tilde \omega_{\text{\text{in}}} t }P^\dagger  + \text{H.c.}\big],
\end{align}
where the rapidly oscillating counter-rotating terms proportional to $e^{\pm i(\omega_{\text{in}}+E) t }$ have been omitted. 
In the following, we will assume that 
$\tilde \epsilon_c=0$, unless otherwise stated.

To evaluate the rate \eqref{xasrate} within the rotating wave approximation, we first note that one can neglect the time derivative  $\partial_t s(t)$  with respect to $\partial_t e^{\pm i\omega_{\text{in}}t}$; only the latter will give a nonzero contribution to  Eq.~\eqref{ixas1} in the limit of a deep core hole. Moreover, when switching to the co-moving frame, with $P \to P e^{-iEt}$, one can again omit fast rotating terms, which would vanish in the integral Eq.~\eqref{ixas1}. Hence, in the rotating wave approximation the absorption rate is given by
\begin{align}
\tilde A_{\text{XAS}}(t)
& = 
g \omega_{\text{in}}
\Big(
is(t)^*e^{i \tilde  \omega_{\text{in}}t}
\langle P(t) \rangle_{g}
+\text{H.c.}\Big)
\\
&=
2g \omega_{\text{in}}\text{Im} \Big(
s(t)^*e^{i \tilde  \omega_{\text{in}}t}
\langle P(t) \rangle_{g}
\Big),
\end{align}
and
\begin{align}
\label{ixas2}
\tilde I_{\text{XAS}} = 
 \lim_{g\to 0} \frac{-2}{g}\,\text{Im}\int_{-\infty}^\infty dt\,  
s(t)^*e^{i \tilde  \omega_{\text{in}}t}
\langle P(t) \rangle_{g}.
\end{align}
In these expressions, the  expectation value is understood in the rotating frame, and the subscript $\langle\cdots\rangle_g$ emphasizes that it is computed for the driven system. 

Starting from Eq.~\eqref{ixas2} [and similarly from Eq.~\eqref{ixas1}] we can also reformulate the XAS intensity in term of response functions: One can express $\langle P(t) \rangle_g$ in linear response using a Kubo formula,
\begin{align}
\label{kubo}
\langle P(t) \rangle_g
=
g\int_{-\infty}^t \!\!d\bar t \,\, \chi_{PP}(t,\bar t)s(\bar t) e^{-i \tilde  \omega_{\text{in}}\bar t} + \mathcal{O}(g^2),
\end{align}
where 
\begin{align}
\chi_{PP}(t,\bar t)
=
-i\big\langle[P(t),P^\dagger(\bar t)]\big\rangle_{g=0}.
\end{align}
Because the core hole is initially filled, and the undriven evolution does not mix core and valence states, only the ordering $PP^\dagger$ contributes to $\chi_{PP}$, i.e., we have 
\begin{align}
\chi_{PP}(t,\bar t)
=
-i\big\langle P( t) P^\dagger(\bar t)\big\rangle_{g=0}
\equiv
\chi_{PP}^>(t,\bar t).
\end{align}
Inserting Eq.~\eqref{kubo} into Eq.~\eqref{ixas2} gives
\begin{align}
\label{ixas3}
I_{\text{XAS}} 
&= 
-2\,\text{Im}
\int_{-\infty}^\infty \!\!\!dt \int_{-\infty}^t \!\!\!d\bar t \,
s(t)^*s(\bar t)e^{i \tilde  \omega_{\text{in}}( t- \bar t)} \chi^>_{PP}(t,\bar t).
\end{align}
With the hermitian symmetry $\chi^>_{PP}(t,\bar t)^*=-\chi^>_{PP}(\bar t,t)$, this can be rewritten as
\begin{align}
\label{ixas8}
I_{\text{XAS}} 
&= 
\int_{-\infty}^\infty\!\! dt d\bar t \,\, i\chi_{PP}^>(t,\bar t) s(t) s(\bar t)^* e^{i \tilde  \omega_{\text{in}}( t-\bar t)}.
\end{align}
If both impurity and core dynamics are diagonal in spin and orbital $\gamma$, the XAS intensity is the sum over the different orbital and spin contributions,
\begin{align}
I_{\text{XAS}} 
&= 
\sum_{\gamma,\sigma} I^{\gamma,\sigma}_{\text{XAS}},
\\
\label{ixas9}
I^{\gamma,\sigma}_{\text{XAS}} 
&=
|p_{\gamma}|^2 \int dt d\bar t \,\, i\chi_{\gamma,\sigma}^>(t,\bar t) s(t) s(\bar t)^* e^{i \tilde  \omega_{\text{in}}( t-\bar t)},
\end{align}
where 
\begin{align}
\chi_{\gamma,\sigma}^>(t,\bar t)=-i\langle P_{\gamma,\sigma}(t)P^\dagger_{\gamma,\sigma}(\bar t)\rangle_{g=0}.
\end{align}

Equations \eqref{ixas2} and \eqref{ixas8} provide two formally equivalent ways to compute $I_{\text{XAS}}$. For a system out of equilibrium, it usually does not increase the computational complexity to add the driving $g\neq 0$ to the simulation. In this case, it can be advantageous to calculate the explicit time evolution \eqref{ixas2} whenever the evaluation of the two-point correlation function $\chi_{PP}(t,\bar t)$  is more involved than the evaluation of the single-time expectation value $\langle P(t) \rangle_g$. (As discussed in Ref.~\onlinecite{Eckstein2021}, computing higher-order correlation functions within the hybridization expansion requires vertex corrections.) However, once  $\chi_{PP}(t,\bar t)$ has been computed at all times, $I_{\text{XAS}}$ can be evaluated from Eq.~\eqref{ixas8}  for all possible probe pulses $s(t)$ and probe frequencies at once, while the use of Eq.~\eqref{ixas2} requires a separate simulation for each probe pulse and frequency. In the present work, we use the explicit evaluation given by Eq.~\eqref{ixas1} in the main text.

For completeness, we also relate the real-time response formula to the conventional spectral representation of the XAS signal (see also Refs.~\onlinecite{Eckstein2021, Zawadzki2020, Chen2019} for similar discussions in the context of RIXS).  In equilibrium, the response function $\chi_{PP}$ is translationally invariant and can be represented in terms of its Fourier transform,
\begin{align}
\chi_{PP}^>(t,\bar t)=\int\frac{d\omega}{2\pi}\chi_{PP}^>(\omega) e^{-i\omega(t-\bar t)}.
\end{align}
The absorption \eqref{ixas8} becomes
\begin{align}
\label{ixas7}
I_{\text{XAS}} 
&= 
\int\frac{d\omega}{2\pi}i\chi_{PP}^>(\omega)
|\tilde s(\omega_{\text{in}}-\omega) |^2.
\end{align}
This is just a convolution of $\chi_{PP}^>(\omega)$ with the square of the Fourier transform of the probe,
\begin{align}
\label{sw}
\tilde s(\omega)
=
\int_{-\infty}^\infty\!\! dt \, s(t)  e^{i \omega t},
\end{align}
which implies the Fourier limitation of the spectral resolution due to a finite probe duration. A finite core hole lifetime corresponds to an additional broadening (see Appendix~\ref{app:tests}). The spectral intensity can be represented in terms of the eigenstates
\begin{align}
\label{lehmann}
i\chi_{PP}^>(\omega)
=
2\pi \sum_{j,f} w_j |\langle f | P^\dagger | j\rangle|^2 \delta(\omega - E_f+E_j),
\end{align}
where $j$ are initial states (without a core hole) with statistical weight $w_j$, and $f$ are final states with a core hole.

\begin{figure}[t]
\begin{center}
\includegraphics[angle=-90, width=0.45\textwidth]{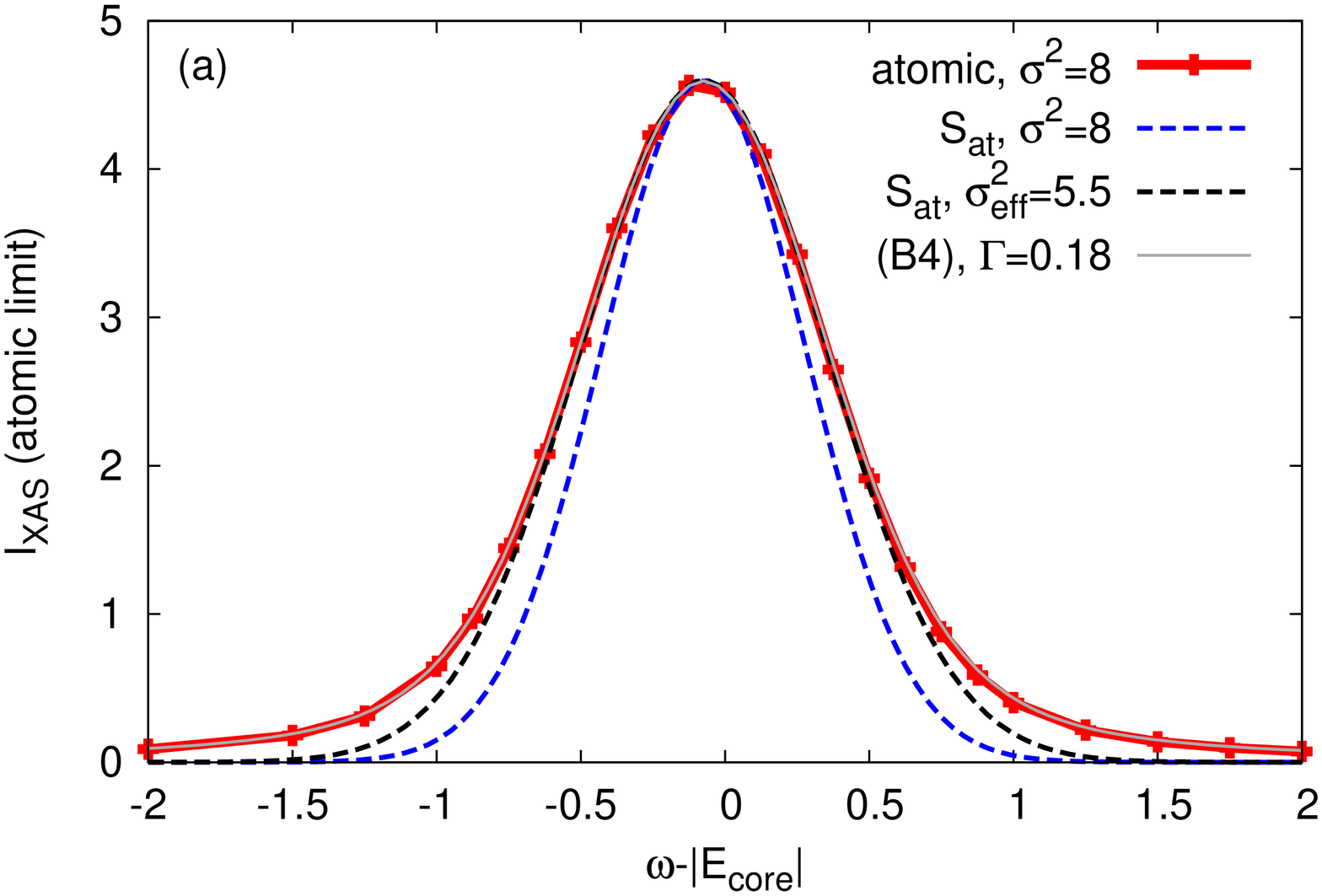}
\includegraphics[angle=-90, width=0.45\textwidth]{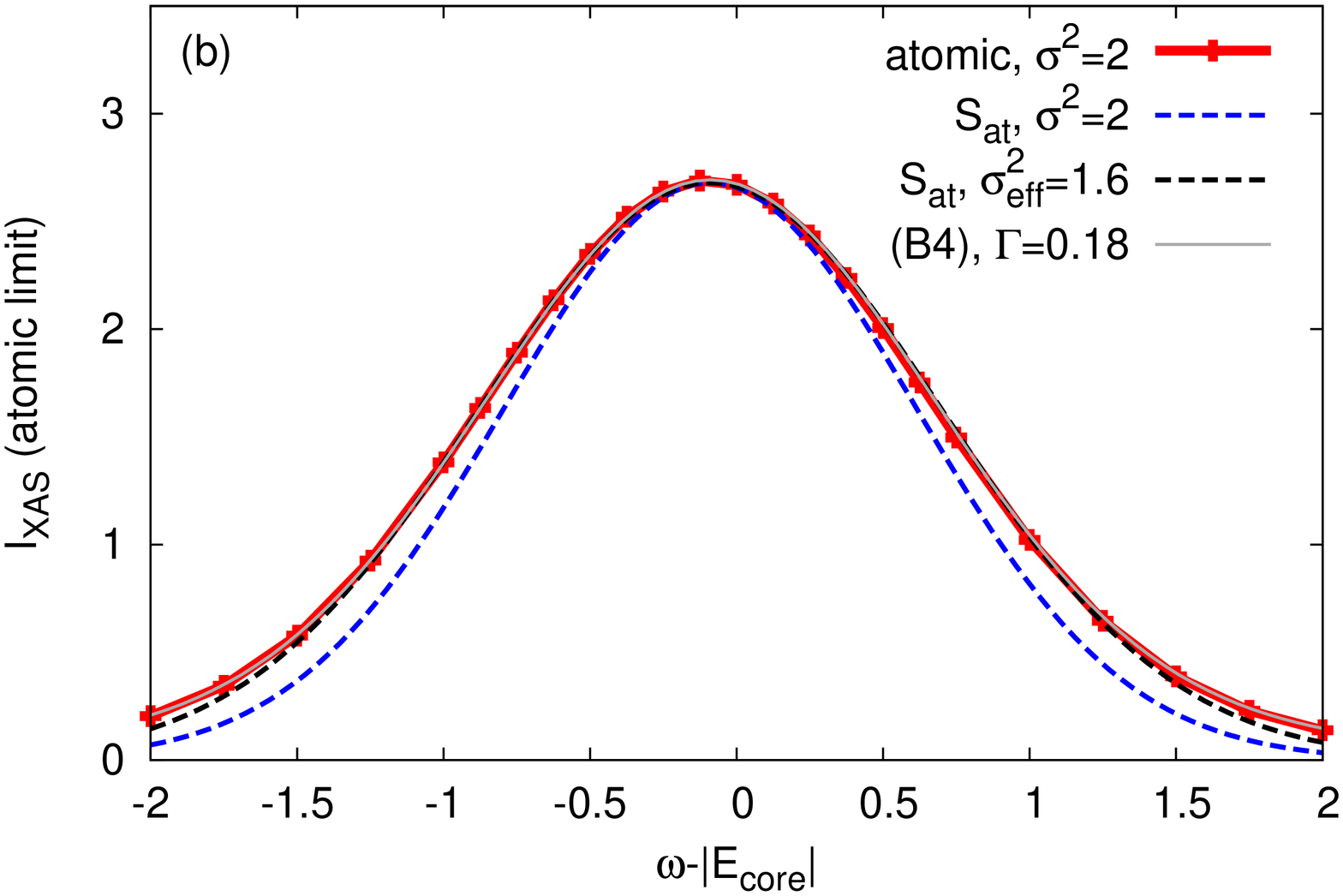}
\caption{
XAS signal of the half-filled atomic system as a function of $\omega-|E_\text{core}|$ for $U=10$, $J=2$, $U_{cd}=6$ (transition $2_h\rightarrow \underline 2_h$).
Panel (a) [(b)] shows data for the probe pulse with $\sigma^2=8$ [$\sigma^2=2$]. 
The solid red lines show $I_\text{XAS}$ for the atomic system with $v_\text{bath}=1$, and the blue dashed lines the power spectrum of the probe pulse (shifted in energy to match the peak position of $I_\text{XAS}$). 
The black dashed lines are fits of the XAS signal with a Gaussian [Eq.~(\ref{atomicline})], with widths $\sigma_\text{eff}^2=5.5$ for the pulse with $\sigma^2=8$ (panel (a)) and $\sigma_\text{eff}^2=1.6$ for the pulse with $\sigma^2=2$ (panel (b)). The convolution of the power spectrum with a Lorentzian with $\Gamma=0.18$ (Eq.~\eqref{convolution}) is indicated by the gray line. 
}
\label{fig_atomic}
\end{center}
\end{figure}

\section{Atomic limit}
\label{app:tests}
The purpose of this appendix is 
(i) to specify the explicit expression for the core bath and to confirm that this provides a suitable way to implement the finite core hole lifetime, 
 and (ii), to analyze the lineshape for the atomic transitions in the present setting (finite duration of the pump and the coupling to the explicit core bath). These lineshapes are then used in the main text to analyze the spectra [c.f.~Eq.~\eqref{atomicline} and the related discussion]. 

In the simulations we consider the half-filled system and choose $E_\text{core}=-20$, as in the main text. We use the impurity model with local Hamiltonian \eqref{ham} to compute the XAS signal from Eq.~\eqref{ixas1}. In the action \eqref{simp}, the valence hybridization term $\mathcal{S}_{dd}$ is set to zero (atomic limit), while the core bath $\mathcal{S}_{cc}$  [Eq.~\eqref{bathcc}] is taken to be the same as in the simulations in the main text: The hybridization $\Delta_c$ in Eq.~\eqref{bathcc} is given in terms of its density of states 
\begin{align}
D_c(\omega) = \frac{v_\text{bath}/W}{[1+e^{\nu(\omega-(E_\text{core}+W/2))}][1+e^{-\nu(\omega-(E_\text{core}-W/2))}]},
\end{align}
with $W=20$, 
a smooth cutoff ($\nu=5$),
and a bath coupling strength $v_\text{bath}=1$. 
The retarded component of $\Delta_c$ is then $\Delta_{c}^{\text{ret}}(t-t')=-i\theta(t-t')\int d\omega D_{c}(\omega)e^{-i\omega(t-t')}$, and the other components are determined accordingly. Because the core bath is well below the chemical potential, it is completely filled ($\Delta_c^< \approx 0$) and will thus simply lead to a filling-in of a core hole state.
 
Without core-hole decay, the atomic XAS spectrum of the half-filled system, measured with the Gaussian probe pulse \eqref{probeulse} features a single line, $I_{\rm XAS} \sim S(\omega-E_\text{ex})$, where $S(\omega)$ is the  
power spectrum of the probe pulse 
i.e.,  the square of the Fourier transform of $s_{\rm pr}(t)$,
\begin{align}
\label{powers}
S(\omega)^2 = e^{-\omega^2\sigma^2/2} \propto |s_{\rm pr}(\omega)|^2.
\end{align}

Figure \ref{fig_atomic}(a) shows the explicit simulation using $\sigma^2=8$, for the atomic limit with the core bath (bold red line), and the power spectrum of the probe pulse (blue dashed line). The latter is rescaled and shifted such that the peak positions match.  
Apparently, the coupling to the core bath leads to an additional broadening.  In the wide band limit, a flat bath density of states implies an approximately exponential core hole decay, corresponding to a Lorentzian broadening in the Fourier transform. Correspondingly, we compare the spectrum of the simulation with the core bath to the lineshape
\begin{align}
L_\Gamma(\omega) \propto \int d\omega_1\, \frac{S(\omega_1)}{\Gamma^2+(\omega+\omega_1)^2},
\label{convolution}
\end{align}
which is obtained by convoluting a Lorentzian with width $\Gamma$ with the power spectrum (thin gray line). We find a good agreement for $\Gamma=0.18$, which indicates that the core bath used in this study approximately reproduces the conventional exponential decay. For the reconstruction of the nonequilibrium spectra in the paper, we use for simplicity  a Gaussian fit to the line [c.f. Eq.~\eqref{atomicline} in the main text], as illustrated by the black dashed line. This approximation catches the lineshape in the peak region, but not the (small) tails. For $\sigma^2=8$, the fit gives the width $\sigma_{\rm eff}^2=5.5$. Figure \ref{fig_atomic}(b) repeats the same analysis for the shorter pulse with $\sigma^2=2$, leading to a best fit with $\sigma_{\rm eff}^2=1.6$.

\bibliography{xas}

\begin{thebibliography}{37}
\expandafter\ifx\csname natexlab\endcsname\relax\def\natexlab#1{#1}\fi
\expandafter\ifx\csname bibnamefont\endcsname\relax
  \def\bibnamefont#1{#1}\fi
\expandafter\ifx\csname bibfnamefont\endcsname\relax
  \def\bibfnamefont#1{#1}\fi
\expandafter\ifx\csname citenamefont\endcsname\relax
  \def\citenamefont#1{#1}\fi
\expandafter\ifx\csname url\endcsname\relax
  \def\url#1{\texttt{#1}}\fi
\expandafter\ifx\csname urlprefix\endcsname\relax\def\urlprefix{URL }\fi
\providecommand{\bibinfo}[2]{#2}
\providecommand{\eprint}[2][]{\url{#2}}

\bibitem[{\citenamefont{van~der Laan et~al.}(1986)\citenamefont{van~der Laan,
  Zaanen, Sawatzky, Karnatak, and Esteva}}]{derLann1986}
\bibinfo{author}{\bibfnamefont{G.}~\bibnamefont{van~der Laan}},
  \bibinfo{author}{\bibfnamefont{J.}~\bibnamefont{Zaanen}},
  \bibinfo{author}{\bibfnamefont{G.~A.} \bibnamefont{Sawatzky}},
  \bibinfo{author}{\bibfnamefont{R.}~\bibnamefont{Karnatak}}, \bibnamefont{and}
  \bibinfo{author}{\bibfnamefont{J.-M.} \bibnamefont{Esteva}},
  \bibinfo{journal}{Phys. Rev. B} \textbf{\bibinfo{volume}{33}},
  \bibinfo{pages}{4253} (\bibinfo{year}{1986}),
  \urlprefix\url{https://link.aps.org/doi/10.1103/PhysRevB.33.4253}.

\bibitem[{\citenamefont{Himpsel et~al.}(1991)\citenamefont{Himpsel, Karlsson,
  McLean, Terminello, de~Groot, Abbate, Fuggle, Yarmoff, Thole, and
  Sawatzky}}]{Himpsel1991}
\bibinfo{author}{\bibfnamefont{F.~J.} \bibnamefont{Himpsel}},
  \bibinfo{author}{\bibfnamefont{U.~O.} \bibnamefont{Karlsson}},
  \bibinfo{author}{\bibfnamefont{A.~B.} \bibnamefont{McLean}},
  \bibinfo{author}{\bibfnamefont{L.~J.} \bibnamefont{Terminello}},
  \bibinfo{author}{\bibfnamefont{F.~M.~F.} \bibnamefont{de~Groot}},
  \bibinfo{author}{\bibfnamefont{M.}~\bibnamefont{Abbate}},
  \bibinfo{author}{\bibfnamefont{J.~C.} \bibnamefont{Fuggle}},
  \bibinfo{author}{\bibfnamefont{J.~A.} \bibnamefont{Yarmoff}},
  \bibinfo{author}{\bibfnamefont{B.~T.} \bibnamefont{Thole}}, \bibnamefont{and}
  \bibinfo{author}{\bibfnamefont{G.~A.} \bibnamefont{Sawatzky}},
  \bibinfo{journal}{Phys. Rev. B} \textbf{\bibinfo{volume}{43}},
  \bibinfo{pages}{6899} (\bibinfo{year}{1991}),
  \urlprefix\url{https://link.aps.org/doi/10.1103/PhysRevB.43.6899}.

\bibitem[{\citenamefont{Haverkort et~al.}(2004)\citenamefont{Haverkort,
  Csiszar, Hu, Altieri, Tanaka, Hsieh, Lin, Chen, Hibma, and
  Tjeng}}]{Haverkort2004}
\bibinfo{author}{\bibfnamefont{M.~W.} \bibnamefont{Haverkort}},
  \bibinfo{author}{\bibfnamefont{S.~I.} \bibnamefont{Csiszar}},
  \bibinfo{author}{\bibfnamefont{Z.}~\bibnamefont{Hu}},
  \bibinfo{author}{\bibfnamefont{S.}~\bibnamefont{Altieri}},
  \bibinfo{author}{\bibfnamefont{A.}~\bibnamefont{Tanaka}},
  \bibinfo{author}{\bibfnamefont{H.~H.} \bibnamefont{Hsieh}},
  \bibinfo{author}{\bibfnamefont{H.-J.} \bibnamefont{Lin}},
  \bibinfo{author}{\bibfnamefont{C.~T.} \bibnamefont{Chen}},
  \bibinfo{author}{\bibfnamefont{T.}~\bibnamefont{Hibma}}, \bibnamefont{and}
  \bibinfo{author}{\bibfnamefont{L.~H.} \bibnamefont{Tjeng}},
  \bibinfo{journal}{Phys. Rev. B} \textbf{\bibinfo{volume}{69}},
  \bibinfo{pages}{020408} (\bibinfo{year}{2004}),
  \urlprefix\url{https://link.aps.org/doi/10.1103/PhysRevB.69.020408}.

\bibitem[{\citenamefont{Basov et~al.}(2017)\citenamefont{Basov, Averitt, and
  Hsieh}}]{Basov2017}
\bibinfo{author}{\bibfnamefont{D.~N.} \bibnamefont{Basov}},
  \bibinfo{author}{\bibfnamefont{R.~D.} \bibnamefont{Averitt}},
  \bibnamefont{and} \bibinfo{author}{\bibfnamefont{D.}~\bibnamefont{Hsieh}},
  \bibinfo{journal}{Nature Materials} \textbf{\bibinfo{volume}{16}},
  \bibinfo{pages}{1077} (\bibinfo{year}{2017}), ISSN \bibinfo{issn}{1476-4660},
  \urlprefix\url{https://doi.org/10.1038/nmat5017}.

\bibitem[{\citenamefont{Giannetti et~al.}(2016)\citenamefont{Giannetti, Capone,
  Fausti, Fabrizio, Parmigiani, and Mihailovic}}]{Claudio2016}
\bibinfo{author}{\bibfnamefont{C.}~\bibnamefont{Giannetti}},
  \bibinfo{author}{\bibfnamefont{M.}~\bibnamefont{Capone}},
  \bibinfo{author}{\bibfnamefont{D.}~\bibnamefont{Fausti}},
  \bibinfo{author}{\bibfnamefont{M.}~\bibnamefont{Fabrizio}},
  \bibinfo{author}{\bibfnamefont{F.}~\bibnamefont{Parmigiani}},
  \bibnamefont{and}
  \bibinfo{author}{\bibfnamefont{D.}~\bibnamefont{Mihailovic}},
  \bibinfo{journal}{Advances in Physics} \textbf{\bibinfo{volume}{65}},
  \bibinfo{pages}{58} (\bibinfo{year}{2016}),
  \urlprefix\url{https://doi.org/10.1080/00018732.2016.1194044}.

\bibitem[{\citenamefont{de~la Torre et~al.}(2021)\citenamefont{de~la Torre,
  Kennes, Claassen, Gerber, McIver, and Sentef}}]{Torre2021}
\bibinfo{author}{\bibfnamefont{A.}~\bibnamefont{de~la Torre}},
  \bibinfo{author}{\bibfnamefont{D.~M.} \bibnamefont{Kennes}},
  \bibinfo{author}{\bibfnamefont{M.}~\bibnamefont{Claassen}},
  \bibinfo{author}{\bibfnamefont{S.}~\bibnamefont{Gerber}},
  \bibinfo{author}{\bibfnamefont{J.~W.} \bibnamefont{McIver}},
  \bibnamefont{and} \bibinfo{author}{\bibfnamefont{M.~A.}
  \bibnamefont{Sentef}}, \bibinfo{journal}{Rev. Mod. Phys.}
  \textbf{\bibinfo{volume}{93}}, \bibinfo{pages}{041002}
  (\bibinfo{year}{2021}),
  \urlprefix\url{https://link.aps.org/doi/10.1103/RevModPhys.93.041002}.

\bibitem[{\citenamefont{Baykusheva et~al.}(2022)\citenamefont{Baykusheva, Jang,
  Husain, Lee, TenHuisen, Zhou, Park, Kim, Kim, Kim et~al.}}]{Baykusheva2022}
\bibinfo{author}{\bibfnamefont{D.~R.} \bibnamefont{Baykusheva}},
  \bibinfo{author}{\bibfnamefont{H.}~\bibnamefont{Jang}},
  \bibinfo{author}{\bibfnamefont{A.~A.} \bibnamefont{Husain}},
  \bibinfo{author}{\bibfnamefont{S.}~\bibnamefont{Lee}},
  \bibinfo{author}{\bibfnamefont{S.~F.~R.} \bibnamefont{TenHuisen}},
  \bibinfo{author}{\bibfnamefont{P.}~\bibnamefont{Zhou}},
  \bibinfo{author}{\bibfnamefont{S.}~\bibnamefont{Park}},
  \bibinfo{author}{\bibfnamefont{H.}~\bibnamefont{Kim}},
  \bibinfo{author}{\bibfnamefont{J.-K.} \bibnamefont{Kim}},
  \bibinfo{author}{\bibfnamefont{H.-D.} \bibnamefont{Kim}},
  \bibnamefont{et~al.}, \bibinfo{journal}{Phys. Rev. X}
  \textbf{\bibinfo{volume}{12}}, \bibinfo{pages}{011013}
  (\bibinfo{year}{2022}),
  \urlprefix\url{https://link.aps.org/doi/10.1103/PhysRevX.12.011013}.

\bibitem[{\citenamefont{Mitrano and Wang}(2020)}]{Mitrano2020}
\bibinfo{author}{\bibfnamefont{M.}~\bibnamefont{Mitrano}} \bibnamefont{and}
  \bibinfo{author}{\bibfnamefont{Y.}~\bibnamefont{Wang}},
  \bibinfo{journal}{Communications Physics} \textbf{\bibinfo{volume}{3}},
  \bibinfo{pages}{184} (\bibinfo{year}{2020}),
  \urlprefix\url{https://doi.org/10.1038/s42005-020-00447-6}.

\bibitem[{\citenamefont{Haverkort et~al.}(2012)\citenamefont{Haverkort,
  Zwierzycki, and Andersen}}]{Haverkort2012}
\bibinfo{author}{\bibfnamefont{M.~W.} \bibnamefont{Haverkort}},
  \bibinfo{author}{\bibfnamefont{M.}~\bibnamefont{Zwierzycki}},
  \bibnamefont{and} \bibinfo{author}{\bibfnamefont{O.~K.}
  \bibnamefont{Andersen}}, \bibinfo{journal}{Phys. Rev. B}
  \textbf{\bibinfo{volume}{85}}, \bibinfo{pages}{165113}
  (\bibinfo{year}{2012}),
  \urlprefix\url{https://link.aps.org/doi/10.1103/PhysRevB.85.165113}.

\bibitem[{\citenamefont{Wegkamp et~al.}(2014)\citenamefont{Wegkamp, Herzog,
  Xian, Gatti, Cudazzo, McGahan, Marvel, Haglund, Rubio, Wolf
  et~al.}}]{Wegkamp2014}
\bibinfo{author}{\bibfnamefont{D.}~\bibnamefont{Wegkamp}},
  \bibinfo{author}{\bibfnamefont{M.}~\bibnamefont{Herzog}},
  \bibinfo{author}{\bibfnamefont{L.}~\bibnamefont{Xian}},
  \bibinfo{author}{\bibfnamefont{M.}~\bibnamefont{Gatti}},
  \bibinfo{author}{\bibfnamefont{P.}~\bibnamefont{Cudazzo}},
  \bibinfo{author}{\bibfnamefont{C.~L.} \bibnamefont{McGahan}},
  \bibinfo{author}{\bibfnamefont{R.~E.} \bibnamefont{Marvel}},
  \bibinfo{author}{\bibfnamefont{R.~F.} \bibnamefont{Haglund}},
  \bibinfo{author}{\bibfnamefont{A.}~\bibnamefont{Rubio}},
  \bibinfo{author}{\bibfnamefont{M.}~\bibnamefont{Wolf}}, \bibnamefont{et~al.},
  \bibinfo{journal}{Phys. Rev. Lett.} \textbf{\bibinfo{volume}{113}},
  \bibinfo{pages}{216401} (\bibinfo{year}{2014}),
  \urlprefix\url{https://link.aps.org/doi/10.1103/PhysRevLett.113.216401}.

\bibitem[{\citenamefont{Gole\ifmmode~\check{z}\else \v{z}\fi{}
  et~al.}(2019)\citenamefont{Gole\ifmmode~\check{z}\else \v{z}\fi{}, Boehnke,
  Eckstein, and Werner}}]{Golez2019}
\bibinfo{author}{\bibfnamefont{D.}~\bibnamefont{Gole\ifmmode~\check{z}\else
  \v{z}\fi{}}}, \bibinfo{author}{\bibfnamefont{L.}~\bibnamefont{Boehnke}},
  \bibinfo{author}{\bibfnamefont{M.}~\bibnamefont{Eckstein}}, \bibnamefont{and}
  \bibinfo{author}{\bibfnamefont{P.}~\bibnamefont{Werner}},
  \bibinfo{journal}{Phys. Rev. B} \textbf{\bibinfo{volume}{100}},
  \bibinfo{pages}{041111} (\bibinfo{year}{2019}),
  \urlprefix\url{https://link.aps.org/doi/10.1103/PhysRevB.100.041111}.

\bibitem[{\citenamefont{Tancogne-Dejean
  et~al.}(2018)\citenamefont{Tancogne-Dejean, Sentef, and
  Rubio}}]{TancogneDejean2018}
\bibinfo{author}{\bibfnamefont{N.}~\bibnamefont{Tancogne-Dejean}},
  \bibinfo{author}{\bibfnamefont{M.~A.} \bibnamefont{Sentef}},
  \bibnamefont{and} \bibinfo{author}{\bibfnamefont{A.}~\bibnamefont{Rubio}},
  \bibinfo{journal}{Phys. Rev. Lett.} \textbf{\bibinfo{volume}{121}},
  \bibinfo{pages}{097402} (\bibinfo{year}{2018}),
  \urlprefix\url{https://link.aps.org/doi/10.1103/PhysRevLett.121.097402}.

\bibitem[{\citenamefont{Strand et~al.}(2017)\citenamefont{Strand,
  Gole\ifmmode~\check{z}\else \v{z}\fi{}, Eckstein, and Werner}}]{Strand2017}
\bibinfo{author}{\bibfnamefont{H.~U.~R.} \bibnamefont{Strand}},
  \bibinfo{author}{\bibfnamefont{D.}~\bibnamefont{Gole\ifmmode~\check{z}\else
  \v{z}\fi{}}}, \bibinfo{author}{\bibfnamefont{M.}~\bibnamefont{Eckstein}},
  \bibnamefont{and} \bibinfo{author}{\bibfnamefont{P.}~\bibnamefont{Werner}},
  \bibinfo{journal}{Phys. Rev. B} \textbf{\bibinfo{volume}{96}},
  \bibinfo{pages}{165104} (\bibinfo{year}{2017}),
  \urlprefix\url{https://link.aps.org/doi/10.1103/PhysRevB.96.165104}.

\bibitem[{\citenamefont{Rinc\'on et~al.}(2018)\citenamefont{Rinc\'on, Dagotto,
  and Feiguin}}]{Rincon2018}
\bibinfo{author}{\bibfnamefont{J.}~\bibnamefont{Rinc\'on}},
  \bibinfo{author}{\bibfnamefont{E.}~\bibnamefont{Dagotto}}, \bibnamefont{and}
  \bibinfo{author}{\bibfnamefont{A.~E.} \bibnamefont{Feiguin}},
  \bibinfo{journal}{Phys. Rev. B} \textbf{\bibinfo{volume}{97}},
  \bibinfo{pages}{235104} (\bibinfo{year}{2018}),
  \urlprefix\url{https://link.aps.org/doi/10.1103/PhysRevB.97.235104}.

\bibitem[{\citenamefont{Petocchi et~al.}(2019)\citenamefont{Petocchi, Beck,
  Ederer, and Werner}}]{Petocchi2019}
\bibinfo{author}{\bibfnamefont{F.}~\bibnamefont{Petocchi}},
  \bibinfo{author}{\bibfnamefont{S.}~\bibnamefont{Beck}},
  \bibinfo{author}{\bibfnamefont{C.}~\bibnamefont{Ederer}}, \bibnamefont{and}
  \bibinfo{author}{\bibfnamefont{P.}~\bibnamefont{Werner}},
  \bibinfo{journal}{Phys. Rev. B} \textbf{\bibinfo{volume}{100}},
  \bibinfo{pages}{075147} (\bibinfo{year}{2019}),
  \urlprefix\url{https://link.aps.org/doi/10.1103/PhysRevB.100.075147}.

\bibitem[{\citenamefont{Gillmeister et~al.}(2020)\citenamefont{Gillmeister,
  Gole{\v z}, Chiang, Bittner, Pavlyukh, Berakdar, Werner, and
  Widdra}}]{Gillmeister2020}
\bibinfo{author}{\bibfnamefont{K.}~\bibnamefont{Gillmeister}},
  \bibinfo{author}{\bibfnamefont{D.}~\bibnamefont{Gole{\v z}}},
  \bibinfo{author}{\bibfnamefont{C.-T.} \bibnamefont{Chiang}},
  \bibinfo{author}{\bibfnamefont{N.}~\bibnamefont{Bittner}},
  \bibinfo{author}{\bibfnamefont{Y.}~\bibnamefont{Pavlyukh}},
  \bibinfo{author}{\bibfnamefont{J.}~\bibnamefont{Berakdar}},
  \bibinfo{author}{\bibfnamefont{P.}~\bibnamefont{Werner}}, \bibnamefont{and}
  \bibinfo{author}{\bibfnamefont{W.}~\bibnamefont{Widdra}},
  \bibinfo{journal}{Nature Communications} \textbf{\bibinfo{volume}{11}},
  \bibinfo{pages}{4095} (\bibinfo{year}{2020}),
  \urlprefix\url{https://doi.org/10.1038/s41467-020-17925-8}.

\bibitem[{\citenamefont{Georges et~al.}(1996)\citenamefont{Georges, Kotliar,
  Krauth, and Rozenberg}}]{Georges1996}
\bibinfo{author}{\bibfnamefont{A.}~\bibnamefont{Georges}},
  \bibinfo{author}{\bibfnamefont{G.}~\bibnamefont{Kotliar}},
  \bibinfo{author}{\bibfnamefont{W.}~\bibnamefont{Krauth}}, \bibnamefont{and}
  \bibinfo{author}{\bibfnamefont{M.~J.} \bibnamefont{Rozenberg}},
  \bibinfo{journal}{Rev. Mod. Phys.} \textbf{\bibinfo{volume}{68}},
  \bibinfo{pages}{13} (\bibinfo{year}{1996}),
  \urlprefix\url{https://link.aps.org/doi/10.1103/RevModPhys.68.13}.

\bibitem[{\citenamefont{Cornaglia and Georges}(2007)}]{Cornaglia2007}
\bibinfo{author}{\bibfnamefont{P.~S.} \bibnamefont{Cornaglia}}
  \bibnamefont{and} \bibinfo{author}{\bibfnamefont{A.}~\bibnamefont{Georges}},
  \bibinfo{journal}{Phys. Rev. B} \textbf{\bibinfo{volume}{75}},
  \bibinfo{pages}{115112} (\bibinfo{year}{2007}),
  \urlprefix\url{https://link.aps.org/doi/10.1103/PhysRevB.75.115112}.

\bibitem[{\citenamefont{Haverkort et~al.}(2014)\citenamefont{Haverkort,
  Sangiovanni, Hansmann, Toschi, Lu, and Macke}}]{Haverkort2014}
\bibinfo{author}{\bibfnamefont{M.~W.} \bibnamefont{Haverkort}},
  \bibinfo{author}{\bibfnamefont{G.}~\bibnamefont{Sangiovanni}},
  \bibinfo{author}{\bibfnamefont{P.}~\bibnamefont{Hansmann}},
  \bibinfo{author}{\bibfnamefont{A.}~\bibnamefont{Toschi}},
  \bibinfo{author}{\bibfnamefont{Y.}~\bibnamefont{Lu}}, \bibnamefont{and}
  \bibinfo{author}{\bibfnamefont{S.}~\bibnamefont{Macke}},
  \bibinfo{journal}{{EPL} (Europhysics Letters)}
  \textbf{\bibinfo{volume}{108}}, \bibinfo{pages}{57004}
  (\bibinfo{year}{2014}),
  \urlprefix\url{https://doi.org/10.1209/0295-5075/108/57004}.

\bibitem[{\citenamefont{L\"uder et~al.}(2017)\citenamefont{L\"uder, Sch\"ott,
  Brena, Haverkort, Thunstr\"om, Eriksson, Sanyal, Di~Marco, and
  Kvashnin}}]{Lueder2017}
\bibinfo{author}{\bibfnamefont{J.}~\bibnamefont{L\"uder}},
  \bibinfo{author}{\bibfnamefont{J.}~\bibnamefont{Sch\"ott}},
  \bibinfo{author}{\bibfnamefont{B.}~\bibnamefont{Brena}},
  \bibinfo{author}{\bibfnamefont{M.~W.} \bibnamefont{Haverkort}},
  \bibinfo{author}{\bibfnamefont{P.}~\bibnamefont{Thunstr\"om}},
  \bibinfo{author}{\bibfnamefont{O.}~\bibnamefont{Eriksson}},
  \bibinfo{author}{\bibfnamefont{B.}~\bibnamefont{Sanyal}},
  \bibinfo{author}{\bibfnamefont{I.}~\bibnamefont{Di~Marco}}, \bibnamefont{and}
  \bibinfo{author}{\bibfnamefont{Y.~O.} \bibnamefont{Kvashnin}},
  \bibinfo{journal}{Phys. Rev. B} \textbf{\bibinfo{volume}{96}},
  \bibinfo{pages}{245131} (\bibinfo{year}{2017}),
  \urlprefix\url{https://link.aps.org/doi/10.1103/PhysRevB.96.245131}.

\bibitem[{\citenamefont{Hariki et~al.}(2018)\citenamefont{Hariki, Winder, and
  Kune\ifmmode~\check{s}\else \v{s}\fi{}}}]{Hariki2018}
\bibinfo{author}{\bibfnamefont{A.}~\bibnamefont{Hariki}},
  \bibinfo{author}{\bibfnamefont{M.}~\bibnamefont{Winder}}, \bibnamefont{and}
  \bibinfo{author}{\bibfnamefont{J.}~\bibnamefont{Kune\ifmmode~\check{s}\else
  \v{s}\fi{}}}, \bibinfo{journal}{Phys. Rev. Lett.}
  \textbf{\bibinfo{volume}{121}}, \bibinfo{pages}{126403}
  (\bibinfo{year}{2018}),
  \urlprefix\url{https://link.aps.org/doi/10.1103/PhysRevLett.121.126403}.

\bibitem[{\citenamefont{Hariki et~al.}(2020)\citenamefont{Hariki, Winder,
  Uozumi, and Kune\ifmmode~\check{s}\else \v{s}\fi{}}}]{Hariki2020}
\bibinfo{author}{\bibfnamefont{A.}~\bibnamefont{Hariki}},
  \bibinfo{author}{\bibfnamefont{M.}~\bibnamefont{Winder}},
  \bibinfo{author}{\bibfnamefont{T.}~\bibnamefont{Uozumi}}, \bibnamefont{and}
  \bibinfo{author}{\bibfnamefont{J.}~\bibnamefont{Kune\ifmmode~\check{s}\else
  \v{s}\fi{}}}, \bibinfo{journal}{Phys. Rev. B} \textbf{\bibinfo{volume}{101}},
  \bibinfo{pages}{115130} (\bibinfo{year}{2020}),
  \urlprefix\url{https://link.aps.org/doi/10.1103/PhysRevB.101.115130}.

\bibitem[{\citenamefont{Higashi et~al.}(2021)\citenamefont{Higashi, Winder,
  Kune\ifmmode~\check{s}\else \v{s}\fi{}, and Hariki}}]{Higashi2021}
\bibinfo{author}{\bibfnamefont{K.}~\bibnamefont{Higashi}},
  \bibinfo{author}{\bibfnamefont{M.}~\bibnamefont{Winder}},
  \bibinfo{author}{\bibfnamefont{J.}~\bibnamefont{Kune\ifmmode~\check{s}\else
  \v{s}\fi{}}}, \bibnamefont{and}
  \bibinfo{author}{\bibfnamefont{A.}~\bibnamefont{Hariki}},
  \bibinfo{journal}{Phys. Rev. X} \textbf{\bibinfo{volume}{11}},
  \bibinfo{pages}{041009} (\bibinfo{year}{2021}),
  \urlprefix\url{https://link.aps.org/doi/10.1103/PhysRevX.11.041009}.

\bibitem[{\citenamefont{Aoki et~al.}(2014)\citenamefont{Aoki, Tsuji, Eckstein,
  Kollar, Oka, and Werner}}]{Aoki2014}
\bibinfo{author}{\bibfnamefont{H.}~\bibnamefont{Aoki}},
  \bibinfo{author}{\bibfnamefont{N.}~\bibnamefont{Tsuji}},
  \bibinfo{author}{\bibfnamefont{M.}~\bibnamefont{Eckstein}},
  \bibinfo{author}{\bibfnamefont{M.}~\bibnamefont{Kollar}},
  \bibinfo{author}{\bibfnamefont{T.}~\bibnamefont{Oka}}, \bibnamefont{and}
  \bibinfo{author}{\bibfnamefont{P.}~\bibnamefont{Werner}},
  \bibinfo{journal}{Rev. Mod. Phys.} \textbf{\bibinfo{volume}{86}},
  \bibinfo{pages}{779} (\bibinfo{year}{2014}),
  \urlprefix\url{https://link.aps.org/doi/10.1103/RevModPhys.86.779}.

\bibitem[{\citenamefont{Eckstein and Werner}(2021)}]{Eckstein2021}
\bibinfo{author}{\bibfnamefont{M.}~\bibnamefont{Eckstein}} \bibnamefont{and}
  \bibinfo{author}{\bibfnamefont{P.}~\bibnamefont{Werner}},
  \bibinfo{journal}{Phys. Rev. B} \textbf{\bibinfo{volume}{103}},
  \bibinfo{pages}{115136} (\bibinfo{year}{2021}),
  \urlprefix\url{https://link.aps.org/doi/10.1103/PhysRevB.103.115136}.

\bibitem[{\citenamefont{Werner et~al.}(2021)\citenamefont{Werner, Johnston, and
  Eckstein}}]{Werner2021a}
\bibinfo{author}{\bibfnamefont{P.}~\bibnamefont{Werner}},
  \bibinfo{author}{\bibfnamefont{S.}~\bibnamefont{Johnston}}, \bibnamefont{and}
  \bibinfo{author}{\bibfnamefont{M.}~\bibnamefont{Eckstein}},
  \bibinfo{journal}{Europhysics Letters} \textbf{\bibinfo{volume}{133}},
  \bibinfo{pages}{57005} (\bibinfo{year}{2021}),
  \urlprefix\url{https://doi.org/10.1209/0295-5075/133/57005}.

\bibitem[{\citenamefont{Werner and Eckstein}(2021)}]{Werner2021b}
\bibinfo{author}{\bibfnamefont{P.}~\bibnamefont{Werner}} \bibnamefont{and}
  \bibinfo{author}{\bibfnamefont{M.}~\bibnamefont{Eckstein}},
  \bibinfo{journal}{Phys. Rev. B} \textbf{\bibinfo{volume}{104}},
  \bibinfo{pages}{085155} (\bibinfo{year}{2021}),
  \urlprefix\url{https://link.aps.org/doi/10.1103/PhysRevB.104.085155}.

\bibitem[{\citenamefont{Chen et~al.}(2019)\citenamefont{Chen, Wang, Jia,
  Moritz, Shvaika, Freericks, and Devereaux}}]{Chen2019}
\bibinfo{author}{\bibfnamefont{Y.}~\bibnamefont{Chen}},
  \bibinfo{author}{\bibfnamefont{Y.}~\bibnamefont{Wang}},
  \bibinfo{author}{\bibfnamefont{C.}~\bibnamefont{Jia}},
  \bibinfo{author}{\bibfnamefont{B.}~\bibnamefont{Moritz}},
  \bibinfo{author}{\bibfnamefont{A.~M.} \bibnamefont{Shvaika}},
  \bibinfo{author}{\bibfnamefont{J.~K.} \bibnamefont{Freericks}},
  \bibnamefont{and} \bibinfo{author}{\bibfnamefont{T.~P.}
  \bibnamefont{Devereaux}}, \bibinfo{journal}{Phys. Rev. B}
  \textbf{\bibinfo{volume}{99}}, \bibinfo{pages}{104306}
  (\bibinfo{year}{2019}),
  \urlprefix\url{https://link.aps.org/doi/10.1103/PhysRevB.99.104306}.

\bibitem[{\citenamefont{Wang et~al.}(2020)\citenamefont{Wang, Chen, Jia,
  Moritz, and Devereaux}}]{Wang2020}
\bibinfo{author}{\bibfnamefont{Y.}~\bibnamefont{Wang}},
  \bibinfo{author}{\bibfnamefont{Y.}~\bibnamefont{Chen}},
  \bibinfo{author}{\bibfnamefont{C.}~\bibnamefont{Jia}},
  \bibinfo{author}{\bibfnamefont{B.}~\bibnamefont{Moritz}}, \bibnamefont{and}
  \bibinfo{author}{\bibfnamefont{T.~P.} \bibnamefont{Devereaux}},
  \bibinfo{journal}{Phys. Rev. B} \textbf{\bibinfo{volume}{101}},
  \bibinfo{pages}{165126} (\bibinfo{year}{2020}),
  \urlprefix\url{https://link.aps.org/doi/10.1103/PhysRevB.101.165126}.

\bibitem[{\citenamefont{Eckstein and Werner}(2010)}]{Eckstein2010}
\bibinfo{author}{\bibfnamefont{M.}~\bibnamefont{Eckstein}} \bibnamefont{and}
  \bibinfo{author}{\bibfnamefont{P.}~\bibnamefont{Werner}},
  \bibinfo{journal}{Phys. Rev. B} \textbf{\bibinfo{volume}{82}},
  \bibinfo{pages}{115115} (\bibinfo{year}{2010}),
  \urlprefix\url{https://link.aps.org/doi/10.1103/PhysRevB.82.115115}.

\bibitem[{\citenamefont{Sch{\"u}ler et~al.}(2020)\citenamefont{Sch{\"u}ler,
  Gole{\v z}, Murakami, Bittner, Herrmann, Strand, Werner, and
  Eckstein}}]{Schueler2020}
\bibinfo{author}{\bibfnamefont{M.}~\bibnamefont{Sch{\"u}ler}},
  \bibinfo{author}{\bibfnamefont{D.}~\bibnamefont{Gole{\v z}}},
  \bibinfo{author}{\bibfnamefont{Y.}~\bibnamefont{Murakami}},
  \bibinfo{author}{\bibfnamefont{N.}~\bibnamefont{Bittner}},
  \bibinfo{author}{\bibfnamefont{A.}~\bibnamefont{Herrmann}},
  \bibinfo{author}{\bibfnamefont{H.~U.} \bibnamefont{Strand}},
  \bibinfo{author}{\bibfnamefont{P.}~\bibnamefont{Werner}}, \bibnamefont{and}
  \bibinfo{author}{\bibfnamefont{M.}~\bibnamefont{Eckstein}},
  \bibinfo{journal}{Computer Physics Communications} p. \bibinfo{pages}{107484}
  (\bibinfo{year}{2020}), ISSN \bibinfo{issn}{0010-4655},
  \urlprefix\url{http://www.sciencedirect.com/science/article/pii/S0010465520302277}.

\bibitem[{\citenamefont{Lysne et~al.}(2020)\citenamefont{Lysne, Murakami, and
  Werner}}]{Lysne2020}
\bibinfo{author}{\bibfnamefont{M.}~\bibnamefont{Lysne}},
  \bibinfo{author}{\bibfnamefont{Y.}~\bibnamefont{Murakami}}, \bibnamefont{and}
  \bibinfo{author}{\bibfnamefont{P.}~\bibnamefont{Werner}},
  \bibinfo{journal}{Phys. Rev. B} \textbf{\bibinfo{volume}{101}},
  \bibinfo{pages}{195139} (\bibinfo{year}{2020}),
  \urlprefix\url{https://link.aps.org/doi/10.1103/PhysRevB.101.195139}.

\bibitem[{\citenamefont{Eckstein and Werner}(2011)}]{Eckstein2011}
\bibinfo{author}{\bibfnamefont{M.}~\bibnamefont{Eckstein}} \bibnamefont{and}
  \bibinfo{author}{\bibfnamefont{P.}~\bibnamefont{Werner}},
  \bibinfo{journal}{Phys. Rev. B} \textbf{\bibinfo{volume}{84}},
  \bibinfo{pages}{035122} (\bibinfo{year}{2011}),
  \urlprefix\url{https://link.aps.org/doi/10.1103/PhysRevB.84.035122}.

\bibitem[{\citenamefont{Ligges et~al.}(2018)\citenamefont{Ligges, Avigo,
  Gole\ifmmode~\check{z}\else \v{z}\fi{}, Strand, Beyazit, Hanff, Diekmann,
  Stojchevska, Kall\"ane, Zhou et~al.}}]{ligges2018}
\bibinfo{author}{\bibfnamefont{M.}~\bibnamefont{Ligges}},
  \bibinfo{author}{\bibfnamefont{I.}~\bibnamefont{Avigo}},
  \bibinfo{author}{\bibfnamefont{D.}~\bibnamefont{Gole\ifmmode~\check{z}\else
  \v{z}\fi{}}}, \bibinfo{author}{\bibfnamefont{H.~U.~R.} \bibnamefont{Strand}},
  \bibinfo{author}{\bibfnamefont{Y.}~\bibnamefont{Beyazit}},
  \bibinfo{author}{\bibfnamefont{K.}~\bibnamefont{Hanff}},
  \bibinfo{author}{\bibfnamefont{F.}~\bibnamefont{Diekmann}},
  \bibinfo{author}{\bibfnamefont{L.}~\bibnamefont{Stojchevska}},
  \bibinfo{author}{\bibfnamefont{M.}~\bibnamefont{Kall\"ane}},
  \bibinfo{author}{\bibfnamefont{P.}~\bibnamefont{Zhou}}, \bibnamefont{et~al.},
  \bibinfo{journal}{Phys. Rev. Lett.} \textbf{\bibinfo{volume}{120}},
  \bibinfo{pages}{166401} (\bibinfo{year}{2018}),
  \urlprefix\url{https://link.aps.org/doi/10.1103/PhysRevLett.120.166401}.

\bibitem[{\citenamefont{Dasari et~al.}(2020)\citenamefont{Dasari, Li, Werner,
  and Eckstein}}]{Dasari2020}
\bibinfo{author}{\bibfnamefont{N.}~\bibnamefont{Dasari}},
  \bibinfo{author}{\bibfnamefont{J.}~\bibnamefont{Li}},
  \bibinfo{author}{\bibfnamefont{P.}~\bibnamefont{Werner}}, \bibnamefont{and}
  \bibinfo{author}{\bibfnamefont{M.}~\bibnamefont{Eckstein}},
  \bibinfo{journal}{Phys. Rev. B} \textbf{\bibinfo{volume}{101}},
  \bibinfo{pages}{161107} (\bibinfo{year}{2020}),
  \urlprefix\url{https://link.aps.org/doi/10.1103/PhysRevB.101.161107}.

\bibitem[{\citenamefont{Marciniak et~al.}(2021)\citenamefont{Marciniak,
  Marcantoni, Giusti, Glerean, Sparapassi, Nova, Cartella, Latini, Valiera,
  Rubio et~al.}}]{Marciniak2021}
\bibinfo{author}{\bibfnamefont{A.}~\bibnamefont{Marciniak}},
  \bibinfo{author}{\bibfnamefont{S.}~\bibnamefont{Marcantoni}},
  \bibinfo{author}{\bibfnamefont{F.}~\bibnamefont{Giusti}},
  \bibinfo{author}{\bibfnamefont{F.}~\bibnamefont{Glerean}},
  \bibinfo{author}{\bibfnamefont{G.}~\bibnamefont{Sparapassi}},
  \bibinfo{author}{\bibfnamefont{T.}~\bibnamefont{Nova}},
  \bibinfo{author}{\bibfnamefont{A.}~\bibnamefont{Cartella}},
  \bibinfo{author}{\bibfnamefont{S.}~\bibnamefont{Latini}},
  \bibinfo{author}{\bibfnamefont{F.}~\bibnamefont{Valiera}},
  \bibinfo{author}{\bibfnamefont{A.}~\bibnamefont{Rubio}},
  \bibnamefont{et~al.}, \bibinfo{journal}{Nature Physics}
  \textbf{\bibinfo{volume}{17}}, \bibinfo{pages}{368} (\bibinfo{year}{2021}),
  \urlprefix\url{https://doi.org/10.1038/s41567-020-01098-8}.

\bibitem[{\citenamefont{Zawadzki et~al.}(2020)\citenamefont{Zawadzki, Nocera,
  and Feiguin}}]{Zawadzki2020}
\bibinfo{author}{\bibfnamefont{K.}~\bibnamefont{Zawadzki}},
  \bibinfo{author}{\bibfnamefont{A.}~\bibnamefont{Nocera}}, \bibnamefont{and}
  \bibinfo{author}{\bibfnamefont{A.~E.} \bibnamefont{Feiguin}},
  \emph{\bibinfo{title}{A time-dependent scattering approach to core-level
  spectroscopies}} (\bibinfo{year}{2020}), \eprint{2002.04142}.

\end{thebibliography}

\end{document}